\documentstyle[aps,prl,twocolumn,epsfig,pstcol,amstex]{revtex}
\newcommand{\etal}{{\em et al.}}                
\newcommand{\dn}[2]{d^{#1}{#2}\,}
\newcommand{\erf}{\operatorname{erf}}
\newcommand{\erfi}{\operatorname{erfi}}

\begin{document}
%
\title{  
   \begin{flushright}
      {\rm DOE/ER/41132-100-INT00}\\
      {\rm UCRL-JC-142857}\\[9mm]
   \end{flushright}
   Observing Non-Gaussian Sources in Heavy-Ion Reactions
}
\author{D.A.~Brown$^{1}$, P. Danielewicz$^{2}$}
\address{
   $^1$University of Washington, Seattle, Washington 98195 and \\
       Lawrence Livermore National Laboratory, Livermore California 94551\\
   $^2$Michigan State University, East Lansing, Michigan 48824\\
}  
\date{\today}
\maketitle
\begin{abstract}
We examine the possibility of extracting non-Gaussian sources from 
two-particle correlations in heavy-ion reactions.  Non-Gaussian sources have
been predicted in a variety of model calculations and may have been seen in
various like-meson pair correlations.  As a tool for this investigation, 
we have developed an improved imaging method that relies on a 
Basis spline expansion of the source functions with an improved implementation 
of constraints.  We examine under what conditions this improved method can
distinguish between Gaussian and non-Gaussian sources.  Finally, we 
investigate pion, kaon, and proton sources from the p-Pb reaction at 
$450$~GeV/nucleon and from the S-Pb reaction at $200$~GeV/nucleon studied by
the NA44 experiment.  Both the pion and kaon sources from the S-Pb correlations 
seem to exhibit a Gaussian core with an extended, non-Gaussian halo.  We also 
find evidence for a scaling of the source widths  with particle mass in the 
sources from the p-Pb reaction.
\end{abstract}
\pacs{PACS numbers: 25.75.-q, 25.75.Gz}

\section{Introduction}

Two-particle correlations have proven to be an important tool for 
experimentally accessing the space-time extent of heavy-ion collisions.  
For like-meson pair correlations (e.g. $\pi$'s and $K$'s), 
the correlation is dominated by the so-called Hanbury-Brown/Twiss (HBT) 
effect (in other words, Bose-Einstein symmetrization of the meson-pair 
wavefunction) and the Coulomb corrected correlations are usually adequately 
parameterized by 
Gaussians~\cite{gkp_71,pratt_90,gelbke_90,pratt_98,wiedemann_99,heinz_99}.  
Since meson final state interactions (FSI) can usually be neglected, 
the Coulomb corrected correlation function becomes 
very nearly the Fourier transform of a source function. 
Thus, a Gaussian correlation corresponds to a Gaussian source
function.  {\em In general, there is no reason to expect the source 
to be Gaussian.}  In fact, non-Gaussian sources may already have been observed 
in data \cite{NA44_K,NA44_pi_1,NA44_pi_2,Brown:1999ka,dbrown_1,dbrown_2}. 

There are several reasons to expect non-Gaussian sources: contributions from 
resonance decays should lead to an exponential 
halo~\cite{sullivan_1,core_halo,Csorgo:1999tn}, effects of 
space-momentum correlations (caused by either 
flow~\cite{pratt_90,panitkin_99_1} or string fragmentation~\cite{lund}) should
lead to a focusing of the source~\cite{panitkin_99_1}, and even simple 
geometry should lead to non-Gaussian sources. 
Experimentally distinguishing between Gaussian and non-Gaussian
sources is difficult and may be complicated by FSI within
the pair.  Recently it was realized that, by applying imaging 
techniques to the correlation data, we may extract the two-particle 
source function directly~\cite{dbrown_1,dbrown_2}.  The imaging has two main 
advantages over the traditional HBT approach:  it is model independent, meaning 
that it may reveal non-Gaussian features in the source, and it can clearly 
separate the effects of the FSI and symmetrization from 
effects due to the source itself.  
This last point requires 
elaboration: imaging extracts source functions that may be directly 
compared using correlations that cannot easily be compared when they arise from
completely different particles.
It is this feature that allows us to compare proton, kaon, and pion sources
from p-Pb and S-Pb reactions from NA44.  
Indeed, a direct comparison of the proton and kaon sources from the p-Pb
reaction suggests a simple scaling of the source
widths that one should expect based on Lund-type string phenomenology in a
fragmenting string.

Extracting the source function, $S_{\bf P} ({\bf r'})$, begins by noting that
$S_{\bf P} ({\bf r'})$ is related to the experimentally measured 
two-particle correlation, $C_{\bf P}({\bf q'})$, through a simple linear 
integral equation~\cite{pratt_90,koonin_77}:
\begin{equation}
	{\cal R}_{\bf P}({\bf q'})\equiv C_{\bf P}({\bf q'}) -1 =
	\int d{\bf r'} \,K({\bf q'}, {\bf r'}) \, S_{\bf P} ({\bf r'}) \,.
	\label{eqn:KP3D_K}
\end{equation}
Thus, ``imaging the source'' means somehow inverting this equation.   
Here primes denote quantities in the pair center of mass (CM) frame. 
Although for imaging purposes it is simplest to write 
Eq.~\eqref{eqn:KP3D_K} in the pair CM, Eq.~\eqref{eqn:KP3D_K} may be
written in any frame as ${\cal R}_{\bf P}({\bf q'})$ is a Lorentz invariant
observable.
In~\eqref{eqn:KP3D_K}, ${\bf P}={\bf p_1}+{\bf p_2}$ is the total momentum of 
the pair in the lab frame.  The ${\bf P}$ subscript indicates the boost from 
the lab to the pair CM frame (${\bf P}/P_0$ is the boost velocity between 
the frames).  The kernel of Eq.~\eqref{eqn:KP3D_K} is 
\begin{equation}
	K({\bf q'},{\bf r'}) = |\Phi_{\bf q'}^{(-)}({\bf r'})|^2-1.
	\label{eqn:kernel}
\end{equation}
The wavefunction, $\Phi^{(-)}$, describes the propagation of the pair from a 
relative separation of ${\bf r'}$ in the pair CM to the detector
with relative momentum ${\bf q'}=\frac{1}{2}({\bf p_1'}-{\bf p_2'})$.  The 
source function itself is the quasi-probability of emitting the pair
a distance of ${\bf r'}$ apart, in the CM frame.  We write the source as a
convolution of Wigner functions, $D({\bf r},t,{\bf P}/2)$:
\begin{equation}\begin{array}{lrl}
   \lefteqn{S_{\bf P}({\bf r'})\equiv}&&\\
      &\displaystyle \int \dn{}{t'}\!\!\int\dn{3}{R}\dn{}{T}\!&
         \!D({\bf R}+{\bf r}/2,T+t/2,{\bf P}/2)\\
      &\times&\! \displaystyle D({\bf R}-{\bf r}/2,T-t/2,{\bf P}/2),
\end{array}\label{eqn:sourcefromrates}\end{equation}
where the variables in the lab frame are understood as functions of the 
variables in the pair CM frame.
Here the Wigner functions are normalized particle emission rates
\begin{equation}
   D({\bf r},t,{\bf p})=\left.\frac{E\dn{7}{N}}
      {\dn{3}{r}\dn{}{t}\dn{3}{p}}\right/ \frac{E\dn{3}{N}}{\dn{3}{p}},
	\label{eqn:emissionrate}
\end{equation}
and may be computed directly from a transport model as discussed in 
\cite{dbrown_1,dbrown_2,panitkin_99_1}.  Due to the time integral 
in~\eqref{eqn:sourcefromrates}, we cannot distinguish whether a given 
${\bf r}'$ is associated with a time separation or a spatial separation.
 

Inverting Eq.~\eqref{eqn:KP3D_K} is generally an ill-posed problem.  This 
means that small fluctuations in the data, even if well within statistical or 
systematic errors, can lead to large changes in the imaged source function.  
Ill-posedness stems from experimental factors (e.g. limited
statistics, finite sized momentum bins, etc.) and the intrinsic resolution 
of the kernel in Eq.~\eqref{eqn:kernel}.  In other fields, this stability 
problem is attacked using a variety of tactics including forcing the source
function to obey known constraints or choosing a representation of the problem
in which the kernel's resolution may be optimized.  Both of these techniques    
were exploited in Ref.~\cite{dbrown_2}.  While the imaging in
Ref.~\cite{dbrown_2} was successful, the restored sources were represented in a
basis that does not exhibit the continuity that we expect to see in the source. 
In this paper, we report a dramatic improvement of the imaging by using a
representation of the source in which we have direct control over the continuity
of the source.  Our choice of representation still allows us to utilize
constraints and to optimize the resolution of the kernel.

This paper is organized as follows.  First, we will set up the problem of
inverting angle-averaged correlations (i.e. expressed in terms of 
$q_{inv}=\sqrt{{\bf q}^2-q_0^2}$) and outline the improved imaging method. 
The details of the imaging method and our representation of the source are 
contained in the appendices.
Next, we apply the imaging method to correlations corresponding to 
Gaussian and non-Gaussian sources.  This will orient us to some of
the issues we will face when examining real data.  Finally, we will confront 
like-pion, like-kaon, and two-proton correlation data from S-Pb collisions at
200~GeV/nucleon from NA44 and p+Pb collisions at 450~GeV/nucleon.  

\section{Statement of the problem}

In this section, we will set up the imaging problem.  To simplify our
discussion, we will consider only angle-averaged correlations and sources. 
First, we will outline the one-dimensional imaging problem and mention some of
our expectations based on experience with Fourier transforms.  Second, we will
outline how we utilize the Basis spline representation.  Finally, we will 
describe our solution using a Bayesian approach to imaging.  The details of the 
Basis spline basis and the imaging itself are included in the appendices.

\subsection{The one-dimensional imaging problem}

The source function and correlation in Eq.~\eqref{eqn:KP3D_K} both may be
expanded in spherical harmonics and the relations between the angular
coefficients are listed in Ref.~\cite{dbrown_1}.  With this expansion, we may
image the individual components of the correlation function and compile a full
three-dimensional imaged source.  When doing this, one must take care in 
interpreting the results: imaging is formulated in the pair CM 
frame as opposed to frame in which we have more intuition, e.g. the lab 
frame.  In this letter, we work with only the first term in the spherical 
expansions, i.e. the angle-averaged source and correlation.  The drawback of
performing a one-dimensional analysis is that the angular information is lost 
and the resulting source function is even more difficult to interpret.

The angle averaged version of Eq.~\eqref{eqn:KP3D_K} is
\begin{equation}
   {\cal R}(q)=C(q)-1=4\pi \int \dn{}{r}r^2 K(q,r) S(r).
\label{eqn:KP1D}
\end{equation}
Here $q=|{\bf q'}|$.  For like pairs in the pair CM frame, $q_0'=0$.  This 
implies that $|{\bf q'}|=q_{inv}=\sqrt{{\bf q}^2-q_0^2}$.

In Eq.~\eqref{eqn:KP1D}, the kernel is simply the kernel in 
Eq.~\eqref{eqn:kernel}, but averaged over the angle between ${\bf q}$ and 
${\bf r}$:
\begin{equation}
   K(q,r)=\frac{1}{2}\int^1_{-1}\dn{}{(\cos\theta_{\bf qr})} K({\bf q},{\bf r}).
\label{eqn:angleavedkernel}
\end{equation}
For identical spin-zero bosons with no FSI, this kernel is 
\begin{equation}
   K(q,r)=\sin{(2qr)}/2qr,
   \label{eqn:noFSIkernel}
\end{equation}
while with FSI it is, 
\begin{equation}
   K(q,r)=\sum_{\ell\;{\rm even}} 
      \frac{\left|g_\ell(r)\right|^2}{(2\ell+1)}-1.
\end{equation}
Here $\ell$ is the orbital angular momentum quantum number. 
Finally, for protons, the spin-averaged kernel is
\begin{equation}
   K(q,r)=\frac{1}{2}\sum_{js\ell\ell'}(2j+1)
      \left(g_{js}^{\ell\ell'}(r)\right)^2-1.
   \label{eqn:protonkernel}
\end{equation}
Here $\ell$ and $\ell'$ are both orbital angular momentum quantum numbers 
and $j$ and $s$ are the total angular momentum and spin quantum numbers.
In the last two cases, $g$ is the relative final-state radial wavefunction.  
For uncorrected meson data, $g$ is the solution of the Klein-Gordon equation 
including the Coulomb potential.  For protons, $g$ is the solution of the 
Schr\"{o}dinger equation using the Coulomb potential and REID93~\cite{sto94} 
nucleon-nucleon potential. 

Given that the identical particle kernels in Eq.~\eqref{eqn:KP3D_K}
or~\eqref{eqn:KP1D} are Fourier transform kernels at large distances, we 
expect our transforms to 
behave like Fourier transforms.  If Eq.~\eqref{eqn:KP1D} were a 
Fourier transform, then by discretizing Eq.~\eqref{eqn:KP1D}, we would be 
converting the imaging  problem into a finite Fourier transform.  In this 
case, the Sampling Theorem tells us how the sizes of the bins and the 
numbers of bins in the appropriate Fourier spaces are related:
\begin{equation}
   \Delta r=\displaystyle\frac{\hbar c \pi}{q_{max}} \;\;\mbox{and}\;\;
   \Delta q=\displaystyle\frac{\hbar c \pi}{r_{max}} 
   \label{eqn:FTguess}
\end{equation}
where $q_{max}=N\Delta q$, $r_{max}=N\Delta r$ and $N$ is the number of bins  
in both  $r$ and $q$ space.

Using these relations, we may get a feeling for 
how structure in the data effect the imaged source.  For example, the low-$q$ 
structure in the data sets the large length scale behavior of the source.  
Conversely, the high-$q$ portion of the data sets the short length scale 
behavior of the source and therefore sets the size of the smallest features 
features we could hope to resolve in the source.  For example, if the 
correlation dies off around a $q\approx 80$~MeV/c, then we should not expect to 
resolve structure smaller than $\Delta r\approx 8$~fm.  Owing to the fact that our
kernel is not a trigonometric function in general, these estimates are
qualitative at best.  Nevertheless, we will often appeal to Fourier theory to
explanations of some of the effects that we see while imaging.

\subsection{The representation of the problem}

In our calculations, we expand the imaged source in a function basis
\begin{equation}
	S(r)=\sum_{i=1}^{N_M} S_i B_i(r),
   \label{eqn:maingenericftnexpansion}
\end{equation}
and, in this basis, the error on the source is
\begin{equation}
	\Delta S(r)=\sqrt{\sum_{i,j=1}^{N_M} \Delta^2S_{ij} B_i(r) B_j(r)}.
\end{equation}
Here $\Delta^2S$ is the covariance matrix of the source coefficients.
Once we average the kernel over momentum bins to account for the
experimental binning,  our inversion problem reduces to the following 
matrix equation:
\begin{equation}
   {\cal R}_i\equiv {\cal R}(q_i) = \sum_{j=1}^{N_M} K_{ij} S_j 
   \label{eqn:forwardprob}
\end{equation}
where the kernel matrix is 
\begin{equation}
   K_{ij}=\frac{4\pi}{\Delta q}\int_{q_i-\Delta q/2}^{q_i+\Delta q/2}\dn{}{q}
          \int_0^\infty \dn{}{r} r^2 K(q,r)B_j(r).
\end{equation}
Here $\Delta q$ is the momentum bin size.
Our source vector is made of the coefficients $S_j$ of the basis function 
representation of the source and our data vector is made of the 
correlation values ${\cal R}_i$.

The function basis which we use to represent our source function must have
several properties: 1) it must be efficient, i.e. requiring few
coefficients to represent a realistic source, 2) it should be continuous,
or at least have continuity as an option, and 3) it should have an adjustable
parameter that we might use to optimize the resolution in a manner analogous to
Ref.~\cite{dbrown_2}.  One obvious possibility is to either use a Laguerre 
expansion~\cite{ftnexpansion} (so that the first term is an exponential fitted 
to the source) or an Edgeworth expansion~\cite{ftnexpansion,kurtosis} (so that 
the leading term is a Gaussian fitted to the source).  The downside of either 
of these choices is that it difficult to adjust the terms in one's expansion 
to maximize the resolution of the inversion.  Furthermore, one could argue 
that if one picks one of these bases and keeps only a few terms in the
expansion, then one biases the inversion to give, for example, only 
Gaussian sources.

We choose to represent the source function in a Basis spline (a.k.a.
b-spline) basis~\cite{de_boor} as this basis has all of the features we
require for a good representation.  Plots of some sample b-splines are 
shown in Fig.~\ref{fig:bsplines} and this basis is detailed in 
Appendix~\ref{append:bsplines}.   B-splines are piecewise 
polynomials and are continuous up to the degree of these polynomials.  The 
$0^{th}$ degree b-spline is the box-spline, making our b-spline expansion a 
natural generalization of the approach in Refs.~\cite{dbrown_1,dbrown_2}.  
Furthermore, in the b-spline basis the concept of the ``edge of a bin'' in the 
box-spline basis is replaced with the concept of a
knot~\cite{de_boor}.  A knot is simply the 
place where the polynomials that make up the b-spline are patched together.  
In the ``optimized discretization''
scheme of Ref.~\cite{dbrown_2}, the edges of the box-splines are varied to
minimize the relative error of the source.  We may generalize this idea to the
b-splines easily by varying the locations of the knots.  We will give 
examples of choosing the knots in subsection~\ref{knots} and we will explain in 
detail how to choose the ``optimal knots'' in Appendix~\ref{append:bsplines}.

\subsection{The reconstruction}

Once we have converted the inversion problem into a matrix inversion by 
choosing a representation of the source, we proceed as in
Refs.~\cite{dbrown_1,dbrown_2,tarantola,parker}
and extract the source.  The details of the the Bayesian approach to imaging
are discussed in the Appendix~\ref{append:imaging} and we summarize the main 
results here.  To obtain the coefficients of the source, we seek the source that 
minimizes the $\chi^2$:
\begin{equation}
   \chi^2=(K\cdot{\bf S}-\boldsymbol{\cal R})^T\cdot
      ({\Delta^2 {\cal R}})^{-1}\cdot(K\cdot{\bf S}-\boldsymbol{\cal R}),
\label{eqn:chisquare}
\end{equation}
where $\Delta^2 {\cal R}$ is the covariance matrix of the correlation data.
The source that does this is:
\begin{equation}
   {\bf S}={\Delta^2 S}\cdot K^T\cdot 
   (\Delta^2 {\cal R})^{-1}\cdot\boldsymbol{\cal R}
   \label{eqn:normeqn}
\end{equation}
The covariance matrix of this source is:
\begin{equation}
   {\Delta^2 S} = (K^T\cdot (\Delta^2 {\cal R})^{-1}\cdot K)^{-1}.
   \label{eqn:covmtx}
\end{equation}
One should note the dependence on the experimental uncertainty 
$\Delta^2 {\cal R}$ in Eqs.~\eqref{eqn:normeqn} and~\eqref{eqn:covmtx}.
In Eq.~\eqref{eqn:normeqn}, the points with the largest error contribute to the
source determination the least.  Also, in Eq.~\eqref{eqn:covmtx} the points who
are most effected by the points with the large error also have the largest
uncertainty.
         
In order to stabilize the inversion, we can take advantage of prior 
information in the form of equality constraints~\cite{tikhonov}.  
An equality constraint is a condition on
the vector of source coefficients that has the generic form 
${\cal C}\cdot{\bf S}={\bf c}$.  One example of such a constraint
is that the source has slope $0$ at the origin.  Such a situation arises if the
normalized particle emission rates, $D$ (c.f Eq.~\eqref{eqn:emissionrate}), 
have a maximum.  In this case we write
\begin{equation}
   S'(r \rightarrow 0)=\sum_{i=1}^{N_M} S_i B_i'(r\rightarrow 0)=0.
\end{equation}
Thus, this case corresponds to ${\cal C}_i=B_i'(r\rightarrow 0)$ and $c=0$.
We can implement this type of constraint by adding a penalty term 
to the $\chi^2$:
\begin{equation}
   \chi^2+\lambda({\cal C}\cdot{\bf S}-{\bf c})^2
\end{equation}
Here $\lambda$ is a trade-off parameter and we may 
vary it in order to emphasize stability in the inversion (by making
$\lambda$ huge) or to emphasize goodness-of-fit (by setting $\lambda$ to zero). 
Such an ability to trade-off stability for goodness-of-fit is discussed in
{\em Numerical Recipes}~\cite{NumRecipes} in detail.  With this 
modification of the $\chi^2$, the imaged source is
\begin{equation}
   {\bf S}={\Delta^2 S}\cdot \left(K^T\cdot (\Delta^2 {\cal R})^{-1}\cdot
    \boldsymbol{\cal R}+\lambda {\cal C}^T\cdot {\bf c}\right),
\end{equation}
and the covariance matrix of source now is
\begin{equation}
   {\Delta^2 S} = \left(K^T\cdot (\Delta^2 {\cal R})^{-1}\cdot K
   +\lambda {\cal C}^T\cdot{\cal C} \right)^{-1}.
\end{equation}

There is another way in which prior information enters into the inversion -- in
the representation we choose for the source.  By using, say, $N_b=2$ in a 
b-spline expansion of the source, we are really assuming that our source and 
its first and second derivatives are all continuous. 

The reader should note that, when we image, we are really finding a probability 
density for the source given the  correlation data rather than the source 
itself.  The set of source coefficients and the covariance matrix 
of the source characterize the height 
and width of this probability distribution.  In the end, we use the
source coefficients as an estimator of the true source. 

\section{Tests of the imaging}

We now explore the imaging in the b-spline basis
by inverting some simple model correlations.  
We consider two model sources, a Gaussian source:
\begin{equation}
   S(r)=\frac{\lambda}{(2\sqrt{\pi}R_G)^3}\exp{(-\frac{r^2}{4R_G^2})}
   \label{eqn:easier-source}
\end{equation}
and a source with a dipole form-factor like shape:
\begin{equation}
   S(r)=\lambda\frac{2}{\pi^2}\frac{R_D}{({r^2}+{4R_D^2})^2}.
   \label{eqn:tougher-source}
\end{equation}
This second source has a roughly Gaussian peak and an extended tail that one
could imagine corresponds to long-time emission of particles.  We chose
this source to facilitate comparison to the experimental results in the next
section.  We pick $R_G=4.5$~fm, $R_D=3.5$~fm and $\lambda=1$.
To generate the correlations, we convolute the source with the proton
kernel in Eq.~\eqref{eqn:protonkernel} and bin the correlation in 6~MeV/c sized
$q$-bins.  
To simulate realistic data, we take the error bars from the real data of
Ref.~\cite{NSCL-pp} and add statistical scatter consistent with these error 
bars.  The data in~\cite{NSCL-pp} are plotted in the same size momentum bins as
in our test, this data is fully corrected for various experimental effects, and
these effects are properly reflected in their estimates for the experimental
uncertainty.
The resulting correlations are shown in Fig.~\ref{fig:pp-corr}. 
In all of our tests, we confine ourselves to proton correlations
because the proton FSI are more important than meson FSI and
place a more demanding test on the imaging.

Our first test is to examine how the quality of the source reconstruction
depends on the number of coefficients in the source expansion.  
In this test we will use only box-splines.  In the second test, we will use 
higher degree b-splines in the reconstruction.  In this test, we will use a 
fixed separation between the knots (equivalent to using equal width 
box-splines).  In the third test, we will demonstrate the use of the 
``optimal knots'' in analogy to the ``optimized discretization'' method of 
Ref.~\cite{dbrown_2}.  In the final test,
we will demonstrate the practical use of equality constraints.

\subsection{Number of coefficients in the source}
Before we begin our imaging tests, we must decide on the size of our imaging
region and set the number of coefficients that we wish to reconstruct.  Naively,
to set $r_{max}$ we might use the Fourier estimates from 
Eqs.~\eqref{eqn:FTguess} giving $r_{max}=103$~fm.  Experience has shown us that
the source is usually lost in statistical noise and is consistent with zero to
within the errors long before this $r_{max}$.  So, we set
$r_{max}=45$~fm, roughly half of what the naive Fourier estimate suggests.  If we
find this is too conservative, we may increase it later.

To set the number of coefficients, we could use the naive Fourier estimate 
again.  Doing so, we find that $\Delta r=4.1$~fm and that we should use 
11 coefficients.  On the other hand, Eq.~\eqref{eqn:forwardprob} suggests 
that the imaging problem is really a problem of simultaneously solving a set 
of linear equations.  Given this, we look at the data and see that there are 
roughly 15-16 points which are different from one and hence contain useful 
information.  This suggests that we try using something like 14 coefficients 
in the source. 

This raises the question of the amount of information in a data set.  If a 
correlation is Gaussian, than one could argue that it contains only two pieces
of information: the height and width of the Gaussian.  On the other hand if
one bins the data, one could argue that there are really $N$
pieces of information corresponding to the number of bins where there is an
apparent signal.  We adopt the second viewpoint, but comment that the
``amount of information'' in a data set is an imprecise concept. 

\begin{figure}
   \epsfig{file=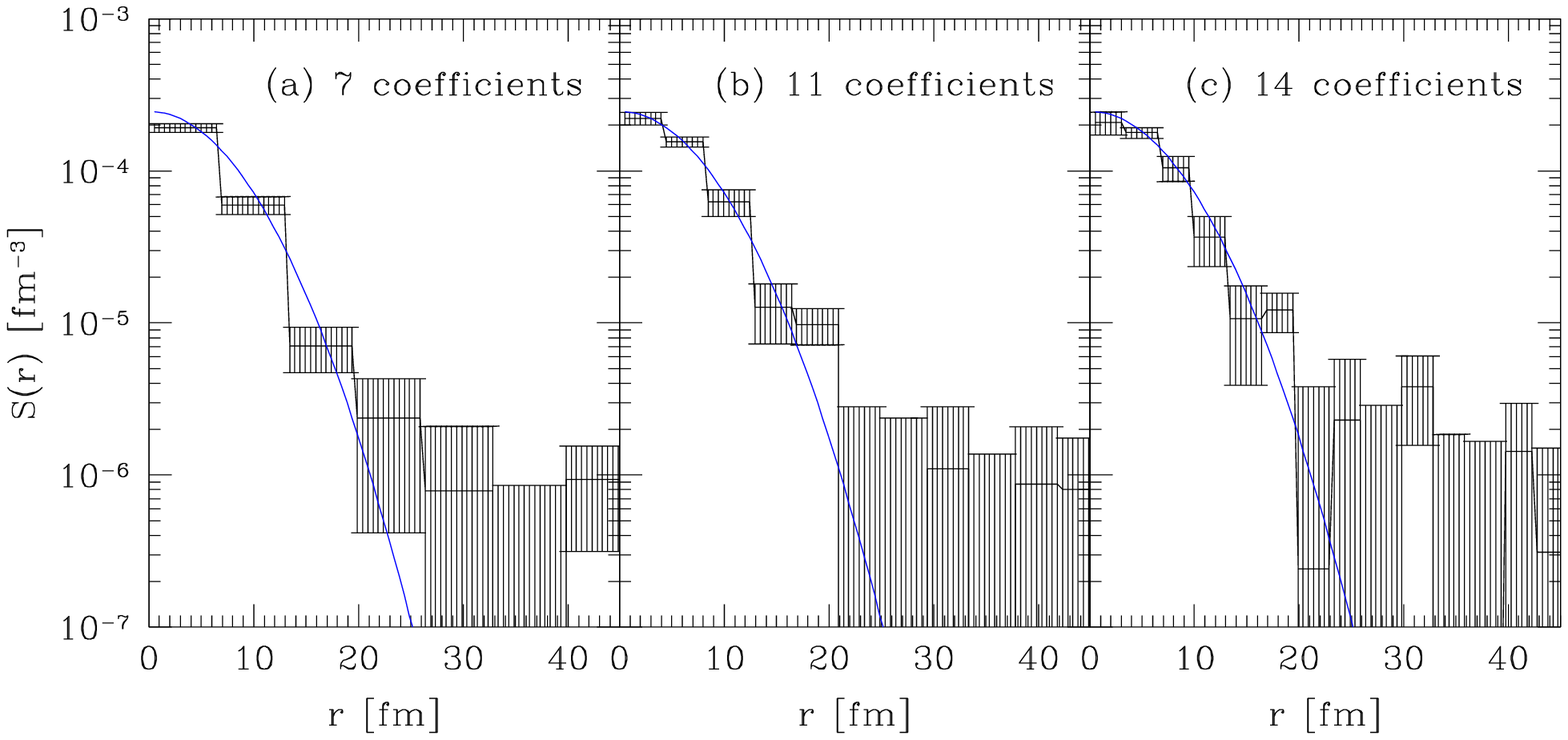,angle=0,width=0.95\textwidth}
   \caption{Reconstructions of the Gaussian source with different 
   numbers of coefficients in the source image.
   In both panels, the model sources are the solid curves and
   the reconstructed source is the curve with the error band.}
   \label{fig:testAa}
\end{figure}

\begin{figure}
   \epsfig{file=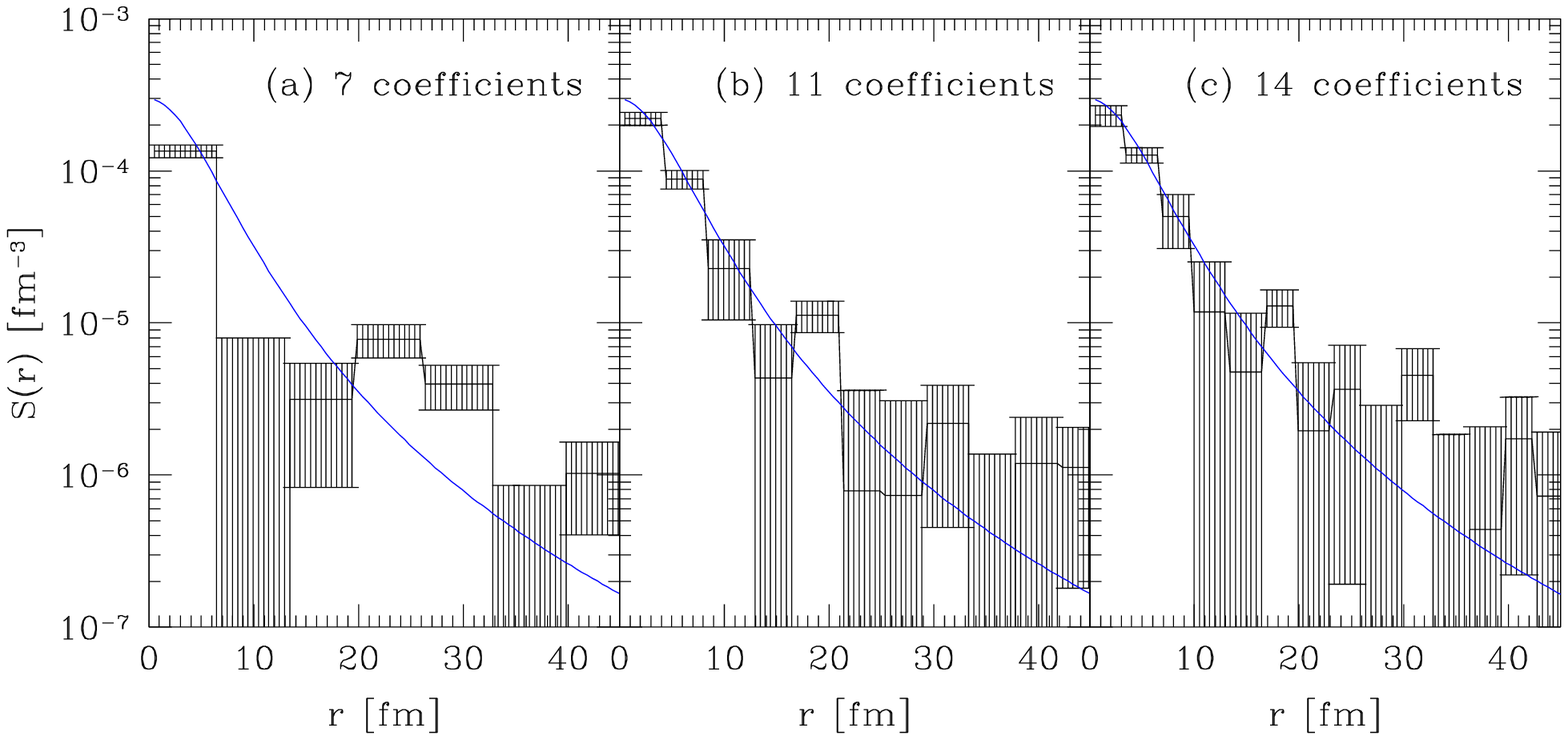,angle=0,width=0.95\textwidth}
   \caption{Reconstructions of the dipole source with different 
   numbers of coefficients in the source image.
   In both panels, the model sources are the solid curves and
   the reconstructed source is curve with the error band.}
   \label{fig:testAb}
\end{figure}

Since it is not clear which number of coefficients we should use, we will try
both.  Additionally, we will image using 7 box-splines so that we may compare 
the results with the higher-degree results of the next section.
In Figs.~\ref{fig:testAa} and~\ref{fig:testAb} we plot the inversions of the
proton correlations in Fig.~\ref{fig:pp-corr}.

First we look at the Gaussian source images in Fig.~\ref{fig:testAa}.
In all three panels, the inversions are reasonable representations of the true
source.  Only by looking at the $\chi^2$ is it clear which image is the
``best:'' for panel (a), $\chi^2=122$, for panel (b), $\chi^2=91$ and for panel
(c),  $\chi^2=76$.  Since there are only $83$ points in the proton correlations,
the inversion with 14 coefficients is ``too good'' and 11 coefficients seems to
be the best choice.  Before moving on to discuss the dipole sources, we mention
that the fluctuations in the imaged sources are not independent.  If they were,
then we would expect that roughly 1/3 of the bins would differ from the true
source by at least a standard deviation.

Now we turn to the dipole sources.  Looking at Fig.~\ref{fig:testAb}, none of
the images are ideal -- all three have large fluctuations that imply that there
is a zero somewhere around $10$~fm.  In these three plots, the $\chi^2$ is not
much of a guide either.  The $\chi^2$'s are 104, 90, and 77 respectively. 
Finally we comment that we cannot tell which images in Figs.~\ref{fig:testAa}
and~\ref{fig:testAb} correspond to Gaussian sources or dipole sources.

We have seen that increasing the number of coefficients in the reconstruction 
helps to reproduce the source, however there is a practical limit to how many 
coefficients we may add.   As the number of source coefficients increases, 
they become less constrained by the data.  At some point there are more source 
coefficients than can be constrained by the data and then the extra coefficients 
only serve to reproduce the high frequency statistical fluctuations in the 
correlation data.  At this point, the imaged source tends to oscillate about 
the true source as we have over-resolved the source.  In general, we can
never tell when we are over-resolving the source as we do not know the true
source.

\subsection{Using the basis spline representation}

We expect the source function to be continuous as it is the convolution of 
two emission rates, yet we represent it by discontinuous box-splines.  
Thus, our first improvement over a simple box-spline representation of the 
source is to use higher degree b-splines.  

\begin{figure}
   \epsfig{file=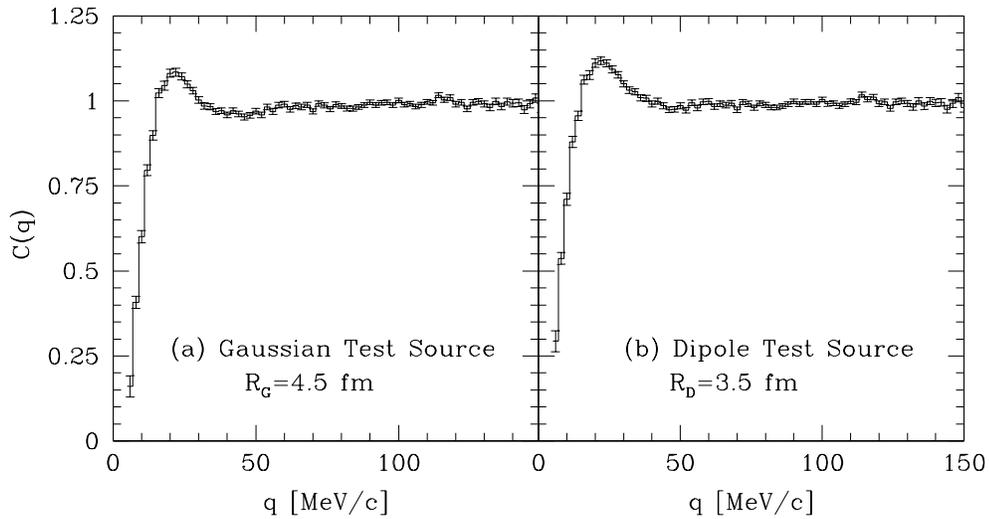,angle=0,width=0.75\textwidth}
   \caption{Model proton correlation corresponding to (a) Gaussian proton source
   function and to (b) Dipole proton source.}
   \label{fig:pp-corr}
\end{figure}

\begin{figure}
   \epsfig{file=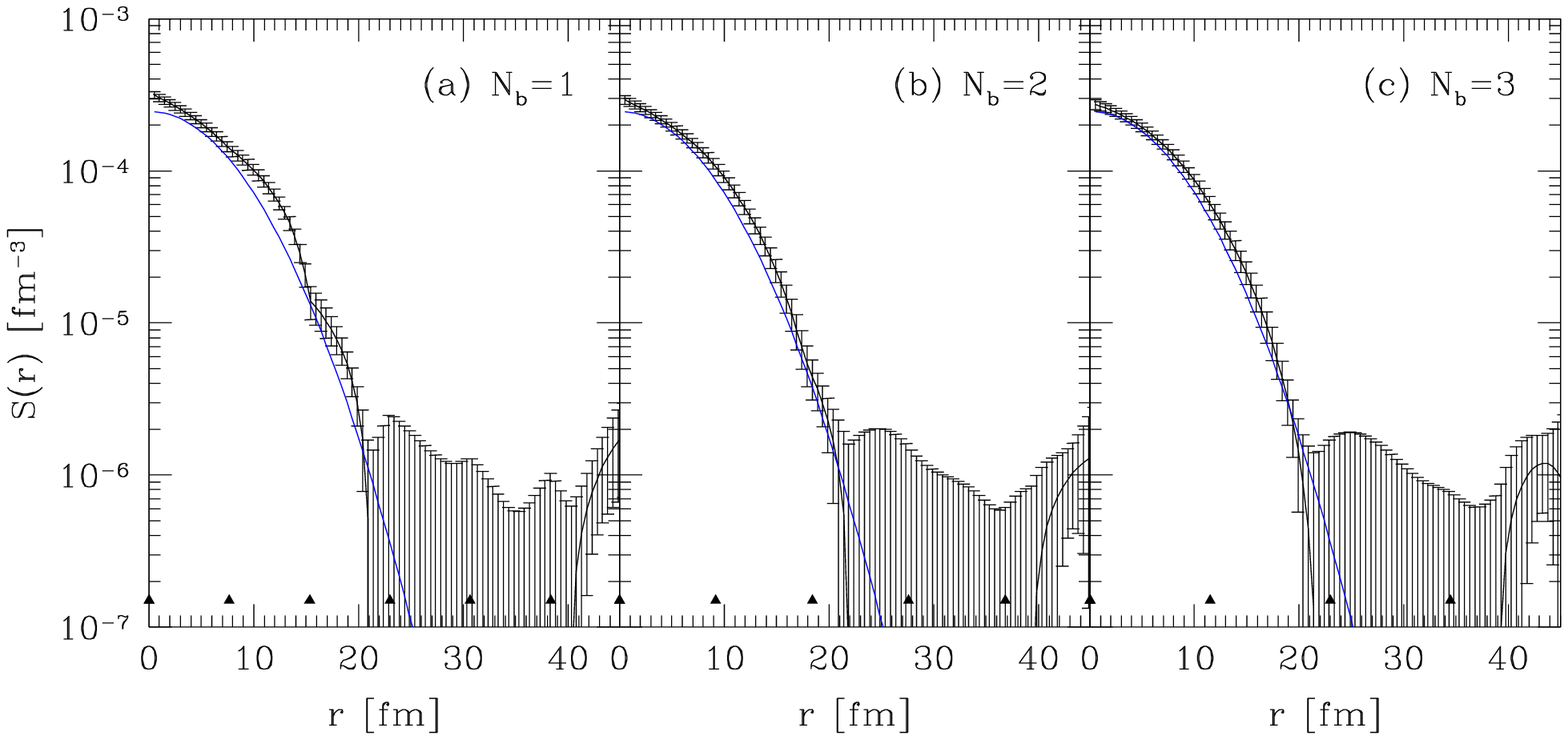,angle=0,width=0.95\textwidth}
   \caption{In all three panels, the true Gaussian source is the solid curve and
   the reconstructed source is given by the points with errors.  The knots in 
   both panels are represented by carets.}
   \label{fig:sou_diff_degrees_4}
\end{figure}

\begin{figure}
   \epsfig{file=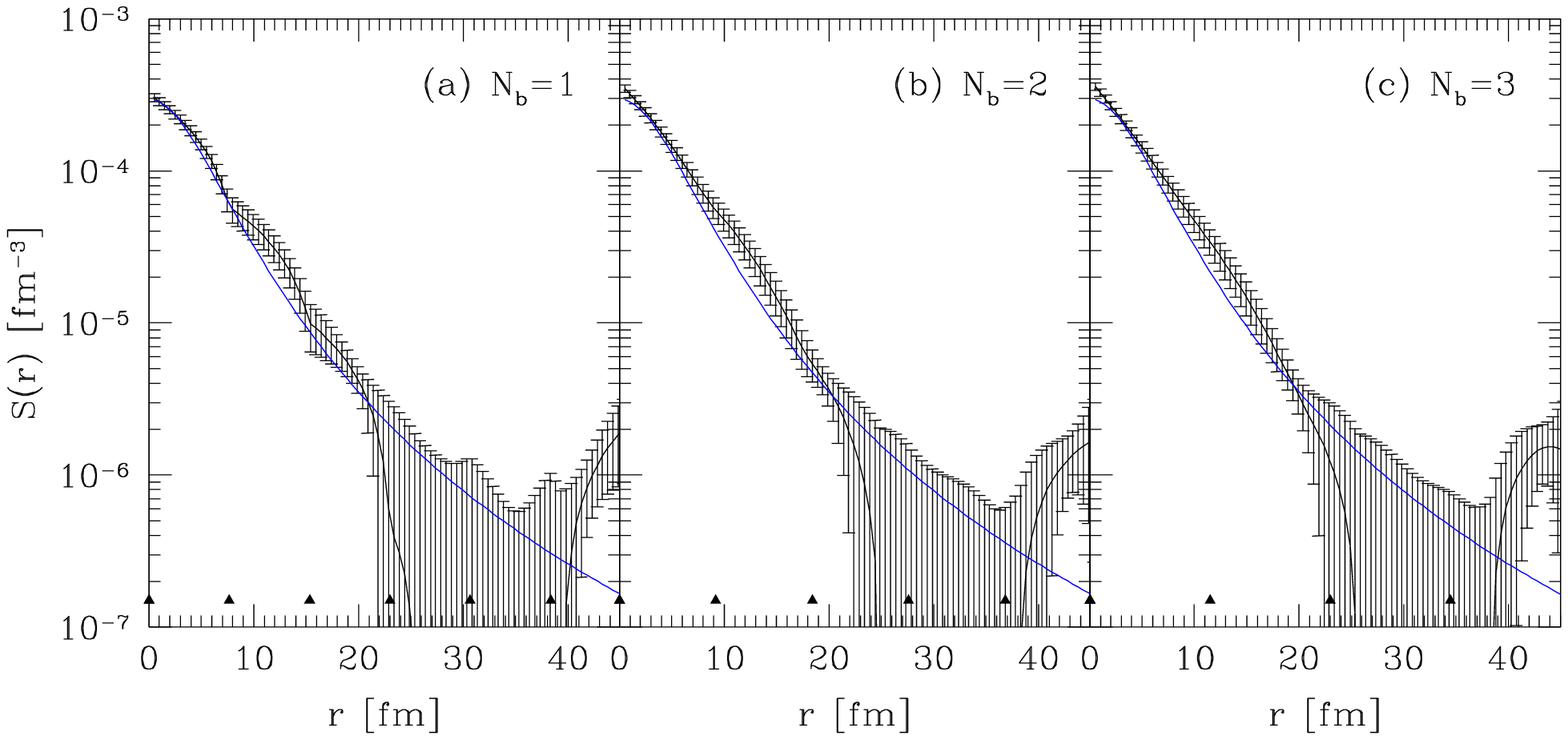,angle=0,width=0.95\textwidth}
   \caption{In three panels, the true dipole source is the solid curve and
   the reconstructed source is given by the points with errors.  The knots in 
   both panels are represented by carets.}
   \label{fig:sou_diff_degrees_5}
\end{figure}

We re-image the proton correlations in Fig.~\ref{fig:pp-corr}.  In 
Figs.~\ref{fig:sou_diff_degrees_4} and~\ref{fig:sou_diff_degrees_5}, we show 
the images obtained using the b-spline expansion with $N_M=7$ coefficients and 
either $1^{st}$, $2^{nd}$ or $3^{rd}$ degree b-splines.  
Our choice of $N_M=7$ is somewhat arbitrary as we do not know how to
extend our Fourier transform based estimates to our b-spline basis.  However,
the fact that the imaging works reasonably well points to the robustness of the 
method, despite the possibly sub-optimal choice of $N_M$.
In all plots, $N_b+1$ knots are placed at the end points of the 
imaging region (i.e. at $r=0$~fm and $r=45$~fm) and the rest of the knots are 
equally spaced between the end points.  

In Fig.~\ref{fig:sou_diff_degrees_4}, the images are fairly accurate 
reconstructions of the source over two orders of magnitude, but the $N_b=3$ 
reconstruction is marginally better, due to the higher degree of continuity.
In all cases, the inversions are better than any of the box-spline results.
The corresponding $\chi^2$'s are $99$ for degree 1 splines, $94$ for degree 2
splines and $93$ for degree 3 splines.
In these plots, the region past $r=17$~fm 
is lost in the noise from the correlation.  We notice that 
all the plots exhibit the same kind of fluctuations seen in the $N_b=0$ images 
in the last section, however they are less noticeable because the b-splines 
are so de-localized.  Finally, we mention that the unphysical rise out past 
40~fm is most likely a result of aliasing.  It is more obvious in these
plots because the last b-spline has a cusp at the edge of the image.

We also image the dipole source in Fig.~\ref{fig:pp-corr}(b).  Using the same
settings as for the Gaussian source, we are able to reproduce the
more complicated behavior of the dipole shaped source over two decades in
source intensity.  More importantly, upon comparing 
Figs.~\ref{fig:sou_diff_degrees_4} and~\ref{fig:sou_diff_degrees_5}, we can 
clearly tell the difference between Gaussian and non-Gaussian source shapes 
on the logarithmic scale.  
This is something we could not do with the box-spline representation of the 
source.
The $\chi^2$ for these reconstructions are 95, 94 and 93 respectively.  We 
comment that the cusp in the very first b-spline helps us represent the 
relatively sharp peak in the dipole source.

\label{aliasing}
In all of the images shown so far, we see an unphysical rise in the source in 
the far right of the images.  This rise is most likely a result of {\em aliasing}.  
Chapter 12 of {\em Numerical Recipes}~\cite{NumRecipes} has a detailed discussion 
of this problem.  Aliasing is a phenomenon that often occurs when 
approximating a Fourier transform over an infinite interval with a finite 
Fourier transform over a finite interval.  Consider a function $f(r)$ and its 
Fourier cosine transform, $F(q)$:
\begin{equation}\begin{array}{rl}
	F(q)&=\displaystyle \int_0^\infty \dn{}{r} f(r) \cos{(qr)}\\
        &\displaystyle \approx\int_0^{r_{max}} \dn{}{r} f(r) \cos{(qr)}.
\end{array}\end{equation}
In the first line of this equation, low frequency structure in $F(q)$ 
corresponds to large distance structure in $f(r)$, which is neglected in the 
second line of this equation.  Now, imagine beginning with $F(q)$ and attempting 
to infer $f(r)$ using a finite $r_{max}$.  What often happens is that, 
whatever strength $f(r)$ should have out past 
$r_{max}$ gets folded into the region $r<r_{max}$.  
In our inversion problem, the integral in 
Eq.~\eqref{eqn:KP3D_K} behaves like a Fourier transform.  
Since statistical fluctuations in the data are artificial high-frequency 
structure, we should not be surprised to see features reminiscent of aliasing 
when we image.  
Based on our experience, adjusting $r_{max}$ or constraining the source at
$r_{max}$ can help cure this problem.
However, the rise at the largest $r$ is usually preceded by a region of 
the image that is consistent with zero so we can easily identify the usable 
part of the image and ignore any artifact due to aliasing.  

\subsection{Choosing the knots}
\label{knots}

For our next refinement, we examine how choosing the knots affects the inversion.
Were the problem of imaging as simple as inverting a Fourier transform, the 
optimal bins in $r$ would be evenly spaced and given by Eq.~\eqref{eqn:FTguess}.  
However, the kernels we are interested in are often distorted by the Coulomb 
repulsion of the pairs as well as other FSI.  Furthermore, some regions of the 
data have large errors and it would be useful if we could combine 
those bins somehow.  Taken together, we must ask whether keeping
equally spaced bins in the source is optimal.  In Ref.~\cite{dbrown_2}, 
we found that we could improve the imaging dramatically by choosing the
size of the bins in $r$ to minimize the error in the source relative to a test
source.  This technique can be generalized to the b-spline basis by
simply varying the knots.
To choose the ``optimal knots'' we proceed as mentioned earlier
and detailed in Appendix~\ref{append:bsplines}.

\begin{figure}
   \epsfig{file=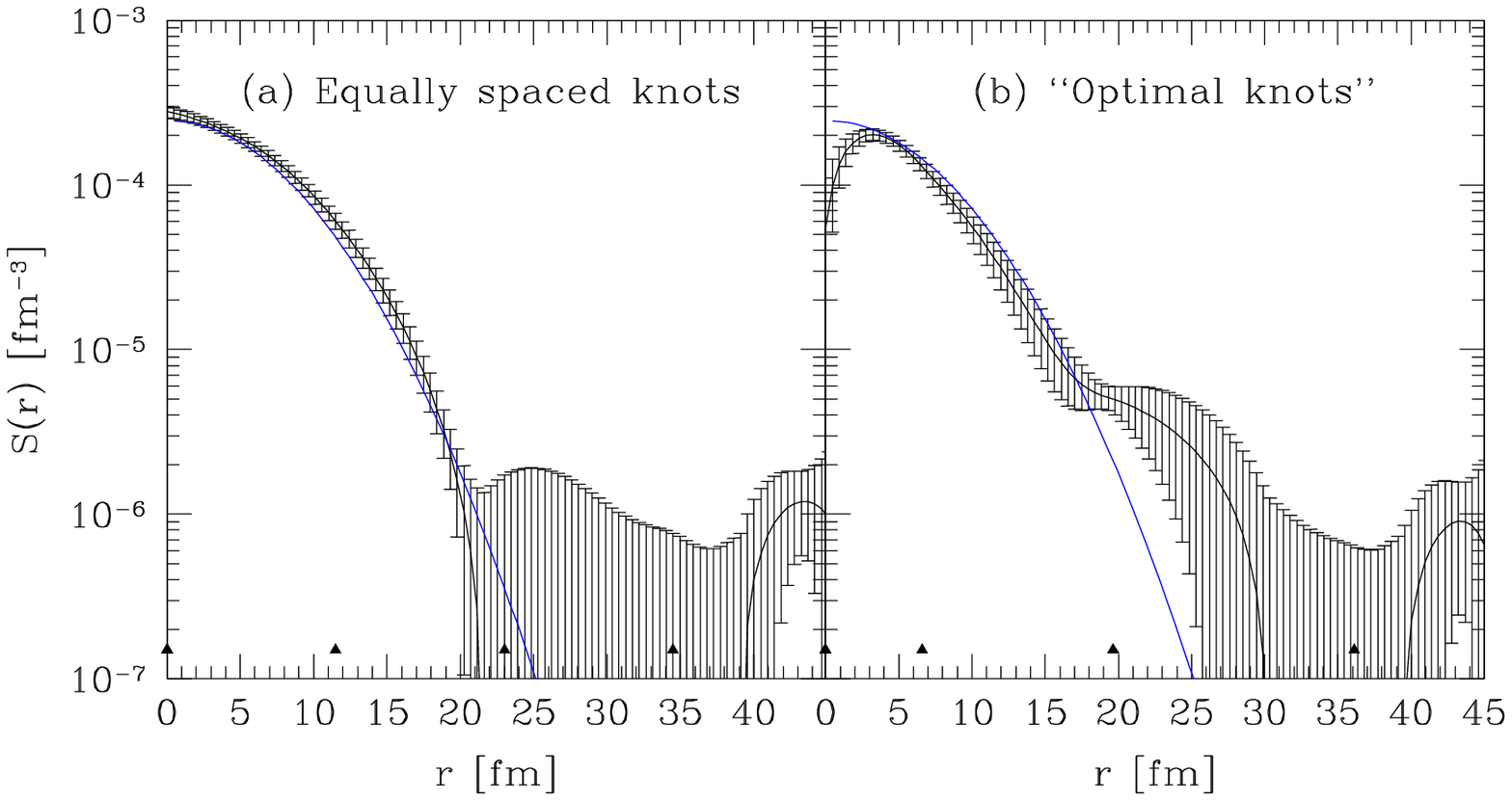,angle=0,width=0.75\textwidth}
   \caption{In both panels, the true Gaussian source is the solid curve 
   and the reconstructed sources is given by the points with errors.  In 
   panel (a), the knots are evenly spaced between the limits of the imaging
   region and in panel (b), the ``optimal knots'' are used.  
   The knots in both panels are represented by carets.  Note that the source
   from panel (c) in Fig.~\ref{fig:sou_diff_degrees_4} is reproduced here as
   panel (a).}
   \label{fig:testC_4}
\end{figure}

For the Gaussian source in Fig.~\ref{fig:testC_4}, we show the inversions 
using $3^{rd}$ degree b-splines using $7$ coefficients for both fixed knots 
(as in Fig.~\ref{fig:sou_diff_degrees_5}(c))
and the ``optimal knots.''  Several things are apparent in this
figure.  First, the fixed width knot reconstruction is markedly better than the
lower-degree b-spline reconstructions in the previous section, simply due to the
higher degree of continuity.  The $\chi^2$ of this reconstruction is 93.
Using the ``optimal knot'' reconstruction, the source is everywhere consistent
with the true source except at the lowest r ($\lesssim 5$~fm) where the source
drops nearly an order of magnitude.  
This drop is unphysical for a source which is the convolution of two single
particle sources, each with one or more maxima.
This drop is due to the close packing of the knots at the lowest r and can be
remedied by lowering the number of coefficients in the reconstruction,
increasing the size of the fit region, or by using equality constraints
(as we will show in the next section).  The $\chi^2$ of this reconstruction 
is 90.

\begin{figure}
   \epsfig{file=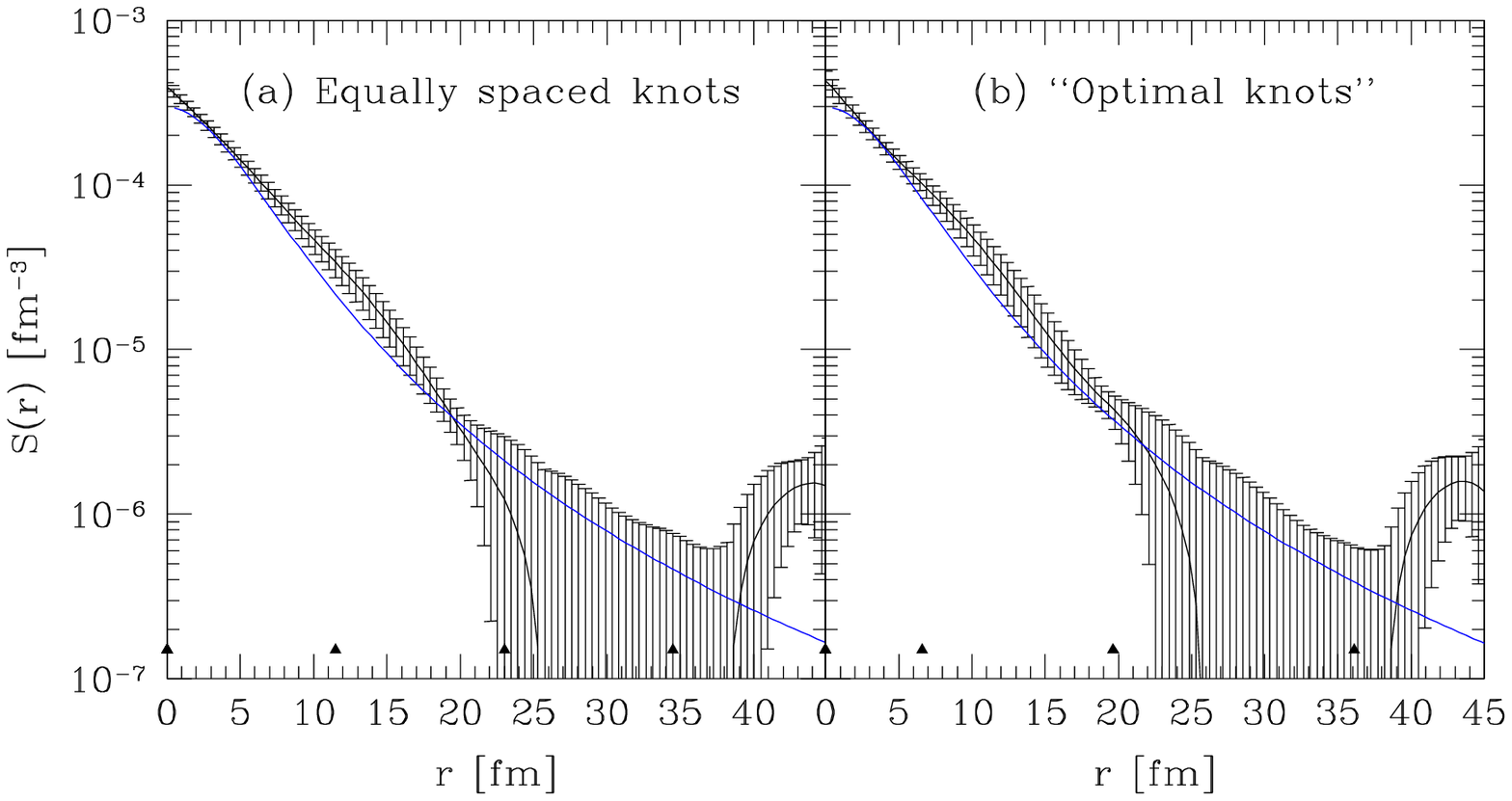,angle=0,width=0.75\textwidth}
   \caption{In both panels, the true dipole source is the solid curve 
   and the reconstructed sources is given by the points with errors.  In 
   panel (a), the knots are evenly spaced between the limits of the imaging
   region and in panel (b), the ``optimal knots'' are used.  
   The knots in both panels are represented by carets.  Note that the source
   from panel (c) in Fig.~\ref{fig:sou_diff_degrees_5} is reproduced here as
   panel (a).}
   \label{fig:testC_5}
\end{figure}

In Fig.~\ref{fig:testC_5}, we show the similar set of inversions for the dipole
shaped source.  Both inversions seem to do a reasonable job of representing the
source, except at the lowest $r$ where the cusp of the first b-spline is a bit
higher than the true source.  The ``optimal knot'' reconstruction is 
marginally better that the equally spaced knot reconstruction as it has a
$\chi^2$ of 92 compared to a $\chi^2$ of 93 for the
equally spaced knot reconstruction.

Given the inconsistent performance of the ``optimal knot'' reconstruction, we
ask ourselves why this refinement does not always help.  To find the optimal 
knots, we move the knots to minimize the error on the source relative to a 
test source (which is the same for all of the inversions).  The error
on the source depends {\em only} on the kernel and the error in the data so the
``optimal knots'' do not know anything about the true source.  If the true
source has interesting structure someplace where we are not sensitive to it, the
``optimal knots'' will be widely spaced there and we will not have the resolution
to see the structure.  Conversely, the ``optimal knots'' may end up giving very
high resolution exactly where we do not need it, witness 
Fig.~\ref{fig:testC_4}(b).  On the other hand, the ``optimal knots'' can give
resolution exactly where we need it, as in Fig.~\ref{fig:testC_5}(b).

%
%
\subsection{Equality Constraints}

Now we come to the final refinement, the use of equality constraints.  As we
have mentioned before, a constraint is a piece of prior information such
as knowing that the first derivative at ${\bf r}=0$
should vanish for a differentiable angle-averaged source, 
$S'(r\rightarrow 0)=0$.  Using constraints amounts to adding information,
so we imagine that we will be able to use more coefficients in the
reconstructions.   This we will see illustrated below.  A list of constraints we 
could use are shown in Table~\ref{table:eqcons}.

\begin{figure}
   \epsfig{file=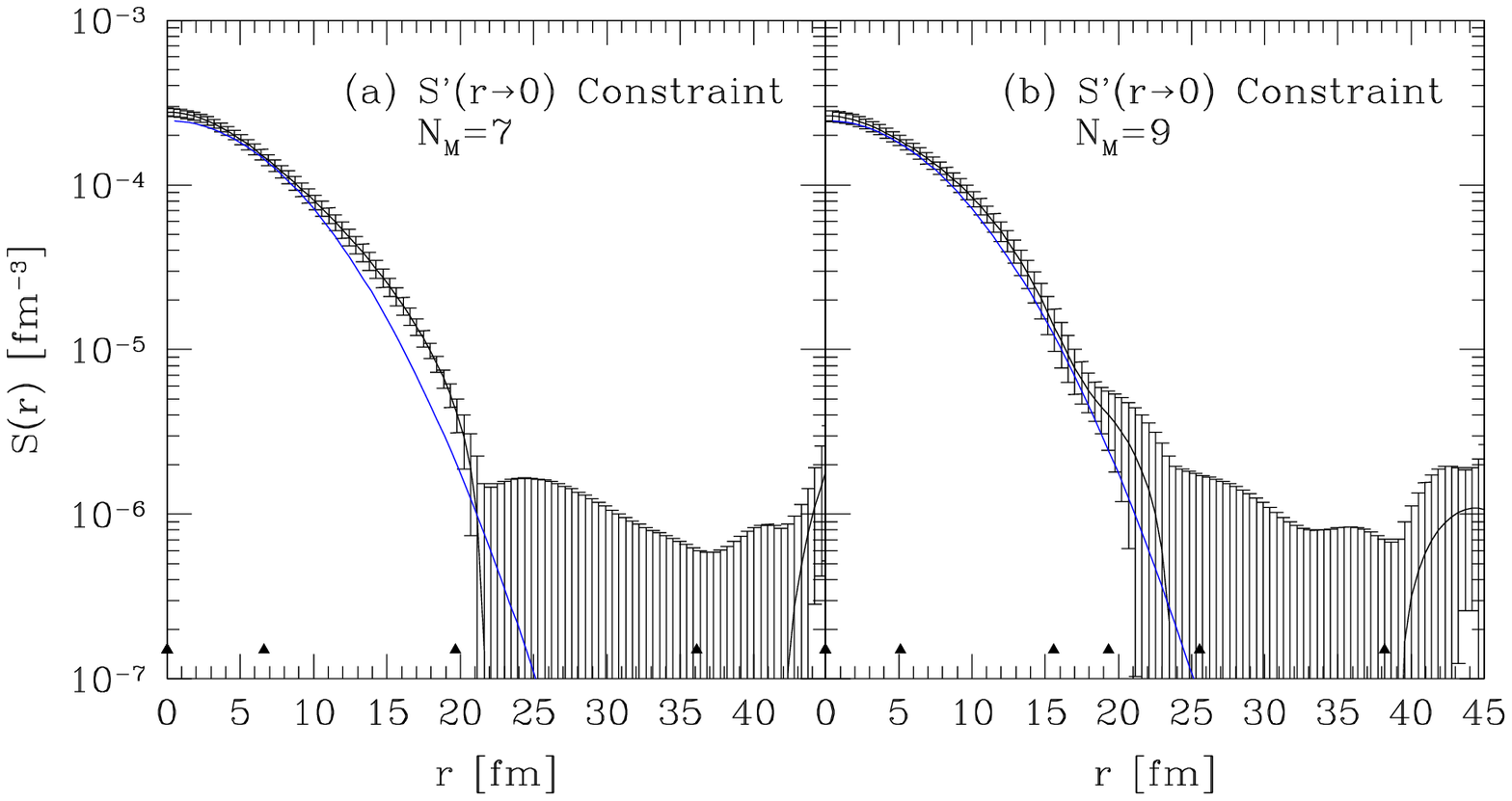,angle=0,width=0.75\textwidth}
   \caption{In both panels, the true Gaussian source is the solid curve and the 
   reconstructed sources are the points with errors.  In both reconstructions,
   the source is constrained to have zero derivative at the origin.  In 
   panel (a), we use 7 coefficients and, in panel (b), 9 coefficients.
   The knots in both panels are represented by carets.}
   \label{fig:testD_4}
\end{figure}

\begin{figure}
   \epsfig{file=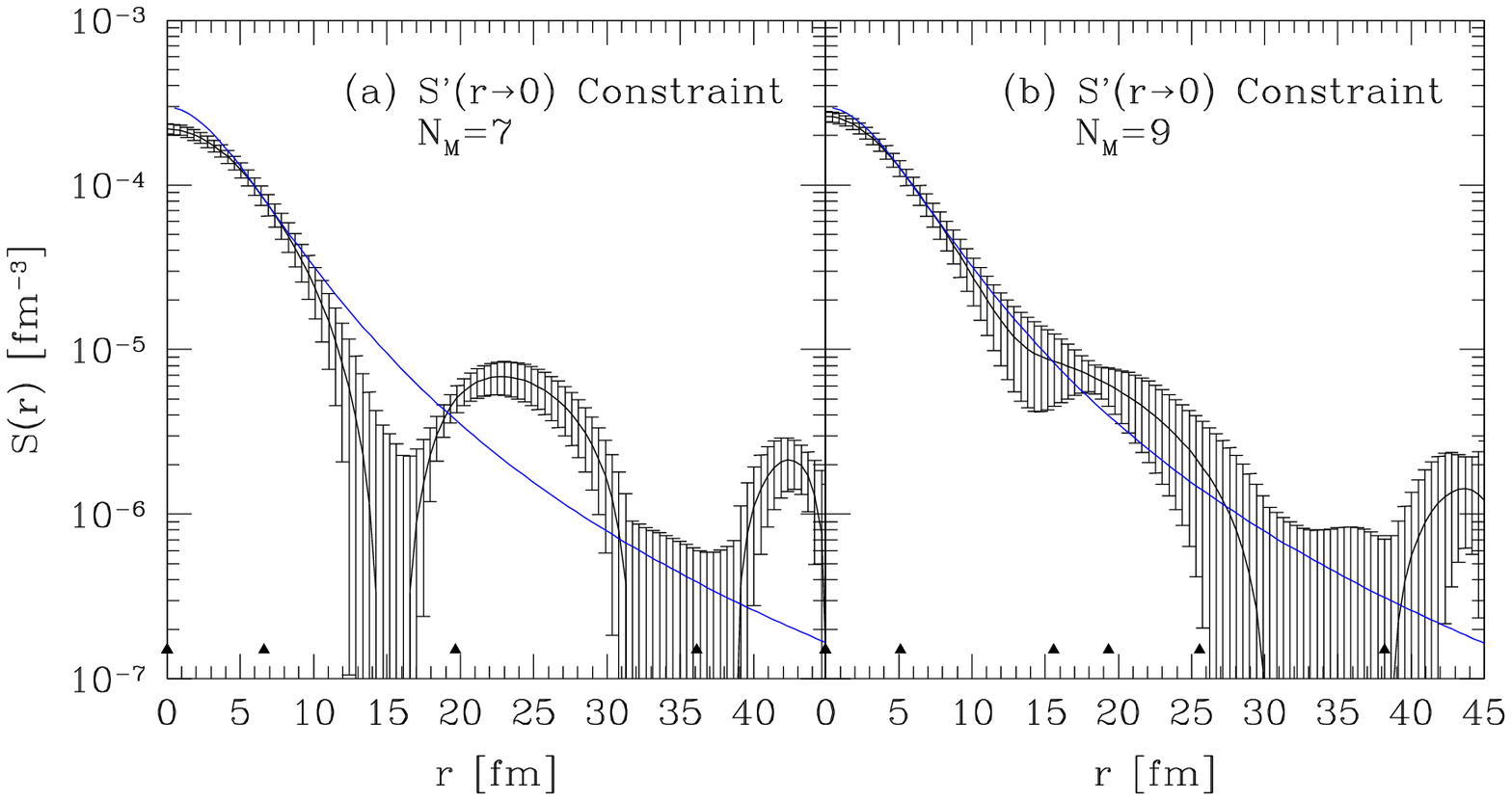,angle=0,width=0.75\textwidth}
   \caption{In both panels, the true dipole source is the solid curve and the 
   reconstructed sources are the points with errors.  In both reconstructions,
   the source is constrained to have zero derivative at the origin.  In 
   panel (a), we use 7 coefficients and, in panel (b), 9 coefficients.
   The knots in both panels are represented by carets.}
   \label{fig:testD_5}
\end{figure}

In Fig.~\ref{fig:testD_4}, we show inversions using the 
$S'(r\rightarrow 0)=0$ constraint for the Gaussian source.  We used
the ``optimal knots,'' $3^{rd}$ degree b-splines and 7 coefficients (in panel
(a)) and 9 coefficients (in panel (b)) in these
inversions.  We see that we have solved the pathological behavior of the imaged
source at low $r$ and the agreement with the true source appears 
good.  Upon examining the $\chi^2$ (107 for panel (a) and 93 for panel (b))
we see that the 7 coefficient source in a lot worse than it appears as it is
routinely higher than the true source in the region from 10-20~fm.  
In Fig.~\ref{fig:testD_5}, we show the results of using the same constraint to
image the dipole-shaped source.  Here we see that, for 7 coefficients (panel(a)),
the quality of the image has gone down considerably: we no longer match the 
height of the peak and we cannot resolve any of the tail.  The $\chi^2$ for this
inversion is an comparatively large 108.  For 9 coefficients (panel(b)), the 
situation is much better.  We now get the peak and can resolve the 
tail.  The $\chi^2$ here is 89.   

We see that this one constraint gave us the ability to add another two points in
the reconstructions without over-resolving the source.  At a
practical level, the first few b-spline coefficients
must be adjusted together in order to satisfy the constraint, in effect leaving
fewer coefficients to fit the data.  Thus, we must add more coefficients if we
want to simultaneously fit the data and satisfy the constraint.
This observation fact leads us to posit a rule of thumb:
``amount of information in the data'' + ``number constraints'' $\geq$
``number of coefficients in expansion.''  Additionally, one should pick the
number of coefficients somewhere near what one would estimate based on the
Fourier estimates discussed earlier. 
    
Finally, by introducing all three refinements of the imaging
(b-splines, ``optimal knots,'' and equality constraints) we are able to
reproduce the height of the source at $r=0$ quite accurately.  The value of the
source at $r=0$ is essential for extracting the space-averaged phase-space
density \cite{dbrown_1,Brown:2000yf}.

\section{Analysis of NA44 Correlations}

\begin{figure}
   \epsfig{file=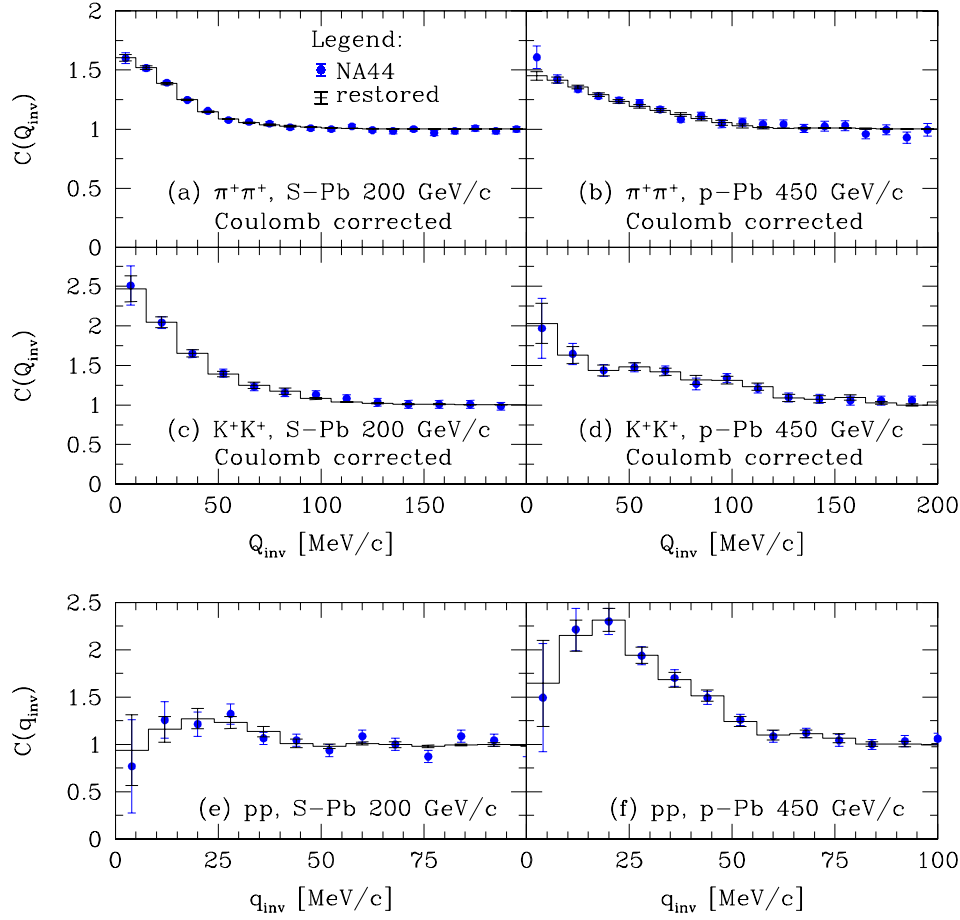,angle=0,width=0.80\textwidth}
   \caption{Pion, kaon and proton correlations for the S-Pb and p-Pb reactions. 
   The points and narrow error bars correspond to the experimentally measured
   correlations.  The histogram and wide error bars correspond to the restored
   correlations from the imaging analysis.  The pion, kaon and proton 
   correlations are from Refs.~\protect\cite{NA44_pi_2},~\protect\cite{NA44_K}, 
   and~\protect\cite{NA44_p}, respectively.}
   \label{fig:NA44-corr}
\end{figure}

Since we can reliably image a source from correlations using the Bayesian
approach to imaging in a b-spline representation, we turn to the analysis 
of NA44 correlations.  In a series of papers
\cite{NA44_K,NA44_pi_1,NA44_pi_2,NA44_p}, NA44 detail
their measurements of angle-averaged pion, kaon, and proton correlations from
p+Pb collisions at 450~AGeV/c and central S+Pb collisions at 200~AGeV/c.
In two of the earlier papers \cite{NA44_K,NA44_pi_1}, they claim to have 
detected non-Gaussian kaon and pion correlations.  To 
bolster their claim, they fit the Coulomb corrected correlations to Gaussians 
and to exponentials.  In particular, they fit to the following 
functional forms:
\begin{equation}
   {\cal R}(Q_{inv})=\lambda \exp{(-Q_{inv}^2R^2_G)}
\end{equation}
implying the Gaussian source of Eq.~\eqref{eqn:easier-source} and 
\begin{equation}
   {\cal R}(Q_{inv})=\lambda \exp{(-2Q_{inv}R_D)}
\end{equation}
implying the source with a dipole-like shape of Eq.~\eqref{eqn:tougher-source}.
Here $Q_{inv}=2q_{inv}=\sqrt{-(p_1-p_2)^2}$, the relative momentum variable
traditionally used in the analysis of meson correlations. The NA44 correlations
that we image are collected in Fig.~\ref{fig:NA44-corr}.

In this section, we first image the NA44 correlations. 
Second, we compare the images to the results of some of NA44's fits.  Next,
we discuss NA44's RQMD simulations of the S-Pb reaction and the implications 
for the source function images.  Finally, we discuss the sources from the
NA44 p-Pb data.

\subsection{Imaging Analysis}

The results of the imaging analysis are presented in Fig.~\ref{fig:NA44-sou}.
As a cross-check, in Fig.~\ref{fig:NA44-corr} we plot the correlations
corresponding to the inverted sources along with the original data.
In these inversions, we used either the noninteracting meson kernel 
in~\eqref{eqn:noFSIkernel} (for the Coulomb corrected pion and kaon 
correlations) or the proton kernel in~\eqref{eqn:protonkernel}.  
Due to the differences in kernels, binning and quality of the various data
sets, each image had to be hand tuned separately.
Since we do not know the true sources that correspond to the data, we used a set
of three criteria to decide when we have a good source:
\begin{enumerate}
   \item Is the image stable -- i.e. when we tweak a parameter (e.g. the
      number of bins, $r_{max}$, number of constraints, etc.) does it 
      change much?
   \item Does the imaged source give a correlation consistent with the
      original?
   \item Is the $\chi^2$ as small as we can make it?
\end{enumerate}
In all cases, we used third degree b-splines.
The parameters of the inversions are collected in 
Table~\ref{table:imag_param}.  Only the p-Pb pp source was imaged
without the $r=0$ smoothness constraint.  We did not use this constraint
because the knot density is too low at low $r$.   Turning on the constraint
widens the peak artificially as the next few b-splines have to be tuned
to get the zero slope at the origin, dramatically raising the $\chi^2$.

When looking at the images, several things are apparent.  First, each of the 
sources from the p-Pb reactions are roughly a factor of two narrower than 
the corresponding sources from the S-Pb reaction.  This is most likely a 
result of the different system sizes.  Second, a comparison of the 
sources from the same reaction reveals that the pion sources are wider than the kaon 
sources and the kaon sources are wider than the proton sources.  Next, it is
apparent that all
six of the sources have main peaks that appear Gaussian.  However, both the 
pion and kaon sources in the S-Pb reaction have significant non-Gaussian tails.  
These tails are most likely not due to aliasing as $r_{max}$ in both plots is 
at 35~fm, but, in order to show all six sources on the same scale we truncated the 
plots at 20~fm.  Unfortunately, this means that we cannot display that the kaon 
and pion sources are consistent with zero in the region from 25-30~fm nor 
can one see the rise that is obviously due to aliasing past the region 30-35~fm 
in the kaon source.  No aliasing is apparent in the pion source or any of the 
proton sources.  In the pion and kaon sources from the p-Pb reaction, aliasing 
is apparent on the far right side of the plots.

\subsection{NA44 Fits}

Following the imaging analysis of the NA44 correlations, it is useful to
compare those results to the various fits performed by NA44.  We have in mind two
sets of fits:  the one-dimensional fits to Gaussians and exponentials in
Ref.~\cite{NA44_K,NA44_pi_1} and the three-dimensional Gaussian fits in
Ref.~\cite{NA44_pi_3}.

\begin{figure}
   \epsfig{file=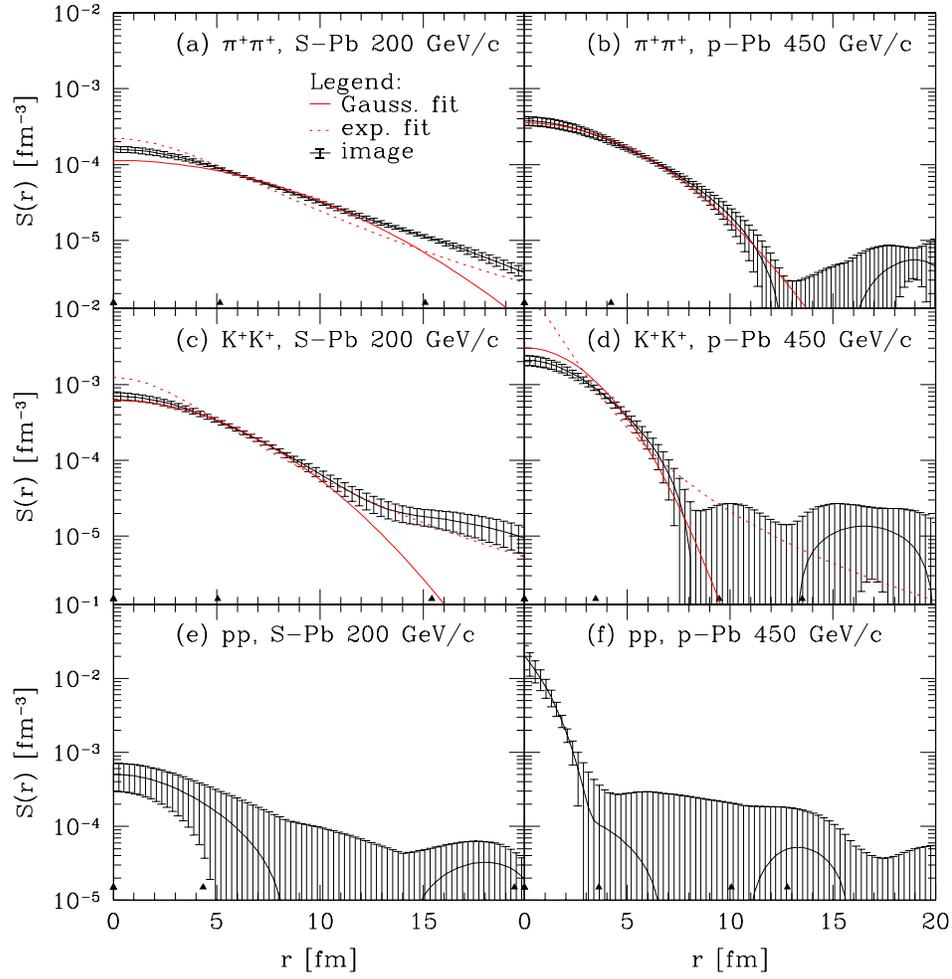,angle=0,width=0.80\textwidth}
   \caption{Sources imaged from the S-Pb and p-Pb reactions.  Where applicable, 
   we have also plotted the Gaussian and dipole-shaped sources corresponding 
   to NA44's fits.}
   \label{fig:NA44-sou}
\end{figure}

We first compare the imaged sources to the results of NA44's one-dimensional
fits.  In addition to the imaged sources, in Fig.~\ref{fig:NA44-sou} we show  
the fits as solid curves (for the Gaussian fits) or dashed curves
(for the exponential fits).  NA44's fit parameters are summarized in 
Tables~\ref{table:gauss_fit_param} and~\ref{table:exp_fit_param}. 
In the S-Pb sources in Fig.~\ref{fig:NA44-sou}, both the kaon and pion
sources seem to be consistent with both the Gaussian and the dipole-shaped
curves in the range from 4~fm out to about 12~fm.  At the lowest $r$ the pion 
image seems to split the difference between the two fits and the kaon image 
is below both fits.  Both sources exhibit long non-Gaussian tails.  In the 
pion source, this tail is higher than the dipole fit and in the kaon source, 
the tail is consistent with the dipole fit. 
For the p-Pb sources in Fig.~\ref{fig:NA44-sou}, both the kaon and
pion sources appear very nearly Gaussian.  At the lowest $r$, the kaon source is
a bit below the Gaussian fit.

\begin{figure}
   \epsfig{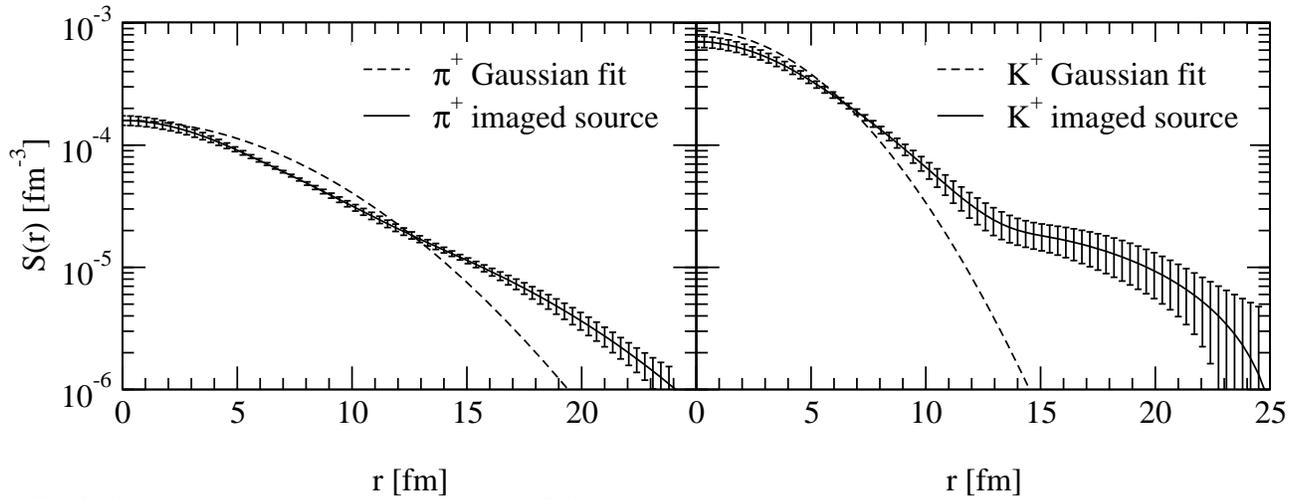}
   \caption{Pion and kaon sources imaged from the S-Pb reactions compared to the
   NA44's angle-averaged three-dimensional sources.}
   \label{fig:NA44-sou3D}
\end{figure}

One might ask if the tails in the sources are simply due to the non-spherical
geometry of the full three-dimensional source.  To answer this question, we
compare the imaged sources to sources corresponding to NA44's three-dimensional
fits in Ref.~\cite{NA44_pi_3}.  These correlations were acquired as follows: 
NA44 first measured like-charged pion and kaon
correlations in the Longitudinal Co-Moving System (LCMS), Coulomb corrected
their data, and then fit their correlations to a Gaussian form.  The form they
fit to is:
\begin{equation}
   C({\bf Q})=1+\lambda \exp{\left( -Q^2_{Ts}R^2_{Ts}
   -Q^2_{To}R^2_{To}-Q^2_{L}R^2_{L}\right)}.
   \label{eqn:cor_gauss3D}
\end{equation}
The results of these fits are shown in Table~\ref{table:gauss3D_fit_param} 
for $\pi^+$'s and $K^+$'s.  Now, as in the one
dimensional case, a three-dimensional Gaussian correlation corresponds to 
a three-dimensional Gaussian source.  For
the correlation in Eq.~\eqref{eqn:cor_gauss3D}, the corresponding source is 
\begin{equation}
   \begin{array}{rl}
      \displaystyle S({\bf r})=
         &\displaystyle \frac{\lambda}{(2\sqrt{\pi})^3R_{Ts}R_{To}R_{L}}\\
         &\displaystyle \times \exp{\left(-\frac{1}{4}\left(
            \frac{r_{Ts}^2}{R^2_{Ts}}+
            \frac{r_{To}^2}{R^2_{To}}+
            \frac{r_{L}^2}{R^2_{L}}\right)\right)}.
   \end{array}
\end{equation}
Looking at the various transverse fit radii obtained by NA44, it seems safe to
set $R_{Ts}=R_{To}\equiv R_{T}$ in this source.  
To convert this three-dimensional source to a one-dimensional source, we only
need to angle-average it, $S(r)=\frac{1}{4\pi}\int \dn{}{\Omega_{\bf r}} S({\bf
r})$, to obtain
\begin{equation}
   \begin{array}{rl}
      S(r)= 
         &\displaystyle\frac{\lambda}{16\pi R^2_{T}R_{L}}
            \exp{\left(-\frac{r^2}{4R_T^2}\right)} \\
         &\times\left\{ 
            \begin{array}{ll}
               \displaystyle\frac{2R_{eff}}{r}
                  \erf{\left(\frac{r}{2R_{eff}}\right)}
                  &\mbox{for $R_T>R_L$}\\
               \displaystyle\frac{2R_{eff}}{r}
                  \erfi{\left(\frac{r}{2R_{eff}}\right)}
                  &\mbox{for $R_T<R_L$}
            \end{array}
         \right.
   \end{array}
\end{equation}
where $R_{eff}=1/\sqrt{|R_L^{-2}-R_T^{-2}|}$ and $\erfi(x)=i\erf(-ix)$.
With this result, we are now able to compare the imaged sources to the
angle-averaged three-dimensional fits from NA44.  This is shown in
Fig.~\ref{fig:NA44-sou3D}.

For these fits, we used $\lambda$ and $R_L$ as given 
Table~\ref{table:gauss3D_fit_param} and $R_T$ was chosen to be the average of 
$R_{Ts}$ and $R_{To}$.  
Examining Fig.~\ref{fig:NA44-sou3D}, it is clear that despite the fact that
NA44's three-dimensional sources correspond to non-Gaussian angle-averaged
sources, they are not able to account for the extended tails we find in the
images.  There is a caveat with this conclusion:
their fits were performed in the LCMS and our 
source is imaged in the pair CM frame.  
Since the two frames coincide only in the limit that $p_T\rightarrow 0$, 
we computed the pion source using the low-$p_T$ data set only.  
Only one data set is available for the kaons.  Nevertheless, it seems 
unlikely that the pion or kaon fit parameters would change  dramatically 
if the fits were performed in the pair CM.

\subsection{Discussion of S-Pb Data}

In addition to the various fits to the correlations, members of the NA44
collaboration performed RQMD 
simulations~\cite{NA44_K,NA44_pi_1,NA44_pi_2,sullivan_1,NA44_p} of the S-Pb 
and p-Pb collisions.  Rather than repeat this work, we will summarize it and 
explain its implications
for the source shape.  In all but the pp case, the simulated 
RQMD correlations compared favorably with the data.  In the pp case, the 
simulations overestimate the height of the correlation peak by roughly 30\%.  

Sullivan \etal~\cite{sullivan_1} explain that the width of the kaon
correlation (corresponding to the width of the imaged kaon sources) is
determined mainly by the size of the kaon production region.
First, kaons are mainly produced directly in the reaction (either from 
fragmenting strings or hadronic reactions) or from the decay of $K^*$ 
resonances.  Now, the reaction zone is roughly the size of the sulfur 
nucleus ($R_{rms}=3.3$~fm) and this defines the size of the Gaussian core.  
The $K^*$'s are also produced in a region of roughly this size.  However, since 
their lifetime is roughly $\tau\approx 4$~fm/c, $K^*$'s do not travel far before 
decaying, giving rise to a non-Gaussian halo surrounding the Gaussian core.  
This halo is neither exponential nor Gaussian, but rather a convolution of a 
Gaussian source with an exponential~\cite{core_halo,Csorgo:1999tn}.
Since the kaons have a much larger mean-free path in nuclear matter
than either pions or nucleons (at least in RQMD), they do not rescatter as the
system evolves.  This means that their last interaction point is very nearly
the size of this production zone.

The pion width in the S-Pb reaction is also described well by the RQMD
calculations.  The initially produced pions (produced mainly from hadronic
reactions, although there is a string component) also have a source width 
set by a combination of the geometrical overlap of the colliding nuclei and 
subsequent dynamics of the system.  From their production until they freeze-out, 
the pions interact strongly with the
system.  Because this evolution involves longitudinal flow, 
there are strong position-momentum correlations in the freeze-out positions of 
the pions.  The longitudinal size of the region where one can find pions with a
low relative momentum is comparable to the transverse size of the system at
freeze-out giving a Gaussian core somewhat larger than the initial system size. 
In addition to this core, the pions also have a large contribution from the 
decay of various resonances, mainly the $\rho$, $\omega$, $\eta'$, and $\eta^0$.
Now the the $\eta$'s are not capable of altering the pion source size or 
shape as the $\eta$'s lifetimes are much too long.  On the other hand, 
both the $\rho$ (whose lifetime is $\approx 2$~fm/c) and the $\omega$ 
(whose lifetime is $\sim 23$~fm/c) both contribute to a non-Gaussian halo in 
a manner analogous to the $K^*$'s above.  Since the $\rho$'s lifetime is smaller
than the $\omega$'s, it contributes to the shorter distance part of the tail and 
the $\omega$ to the longer distance part.

In the case of the protons in the S-Pb reaction, the size of the source is
mainly set by the size of the region where we find protons with low-relative
momentum at freeze-out.  Looking at the plots of RQMD correlations 
in~\cite{NA44_p}, the RQMD simulations are somehow unable to reproduce the 
sizes of these regions.  Since the size of this region is a direct function of 
the geometry of the source and 
the space-momentum correlations in the source (these correlations are  
caused primarily by flow), RQMD's failure here is somewhat of a mystery. 
Unfortunately, our image is not detailed enough to give much
of a clue where the discrepancy arises.  A higher resolution measurement of 
this correlation would help immensely. 

\subsection{Discussion of p-Pb Data}

Unlike the S-Pb reaction, RQMD is able to describe all of NA44's measured 
p-Pb correlations quite well~\cite{NA44_K,NA44_pi_2,NA44_p}.  Because the 
p-Pb system so is small, the reaction is most likely dominated by the formation 
of a few color strings and/or ropes.  RQMD uses the 
Lund string model to model string formation and 
fragmentation~\cite{Sorge:1995}, so to understand what is happening in the 
sources it is useful to understand some of the features of this type of model.  
In the Lund model, momentum and spatial rapidities are tightly correlated.  We
will show that this correlation leads to an approximate scaling of the $K$ and 
$p$ source radii with mass:
\begin{equation}
	m_K R_K\approx m_p R_p.
	\label{eqn:scaling}
\end{equation}
As we will point out, this scaling is born out by the images.
We mention that the pion source {\em should not} 
follow this scaling because a large fraction 
of the pions result from 
resonance decays which distort the shape of the pion source.

\begin{figure}
	\epsfig{file=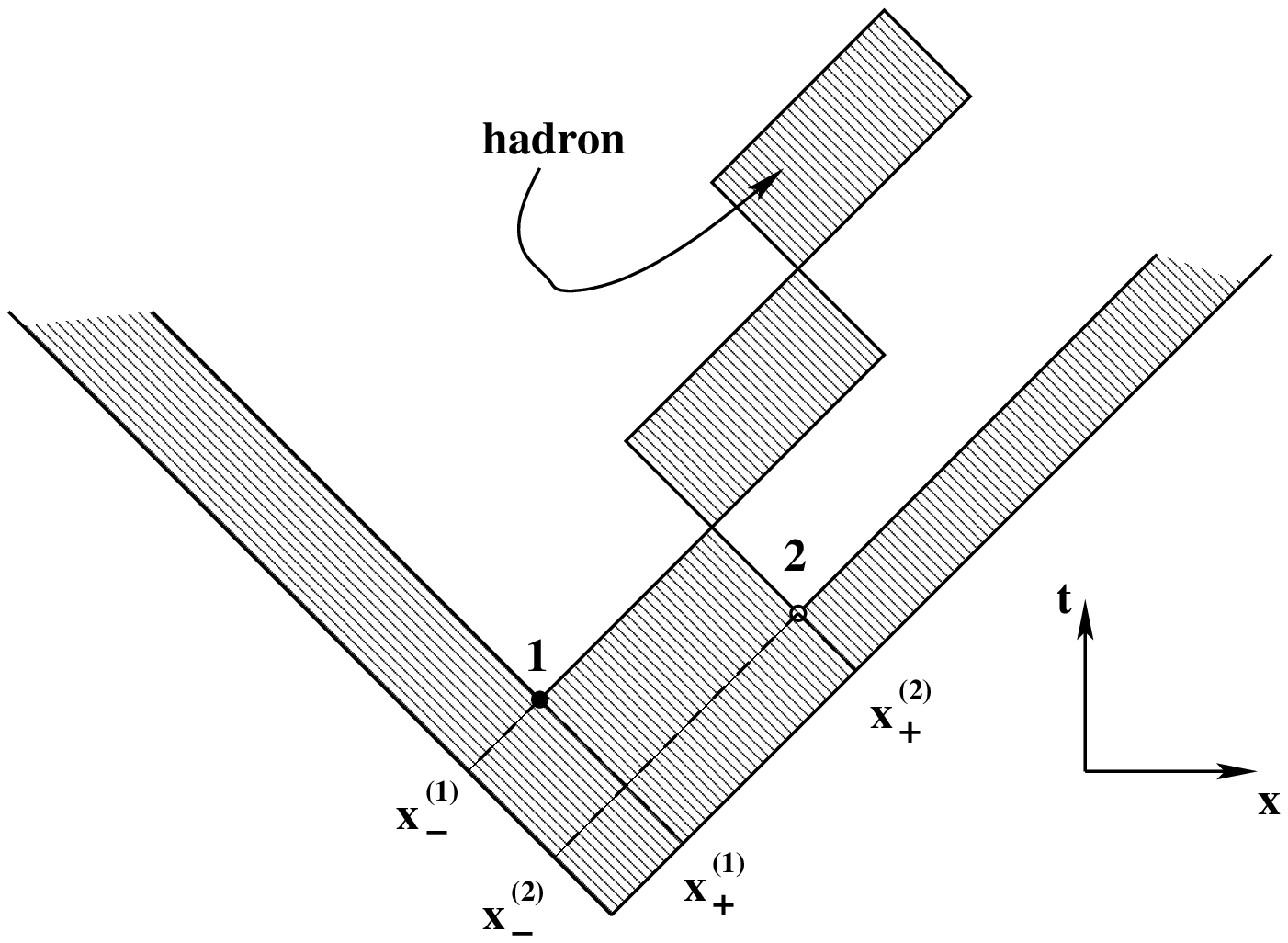,angle=0,width=0.60\textwidth}
	\caption{String fragmentation in the Lund string model.  The hatched areas 
	indicate regions of non-vanishing color field as well as the space-time
	region swept out by the oscillating strings.  Points 1 and 2 indicate 
	the space-time point where the hadron of interest breaks off of the 
	main string.}
	\label{fig:string}
\end{figure}

In a Lund-type string model~\cite{lund:1983}, a meson is viewed as a $q\bar{q}$ 
pair attached by a string and this pair oscillates back and forth in the linear 
confining potential of the sting.  This type of model can also describe baryons 
if one replaces one of the quarks at the end of the string with a di-quark.  
In this discussion, we ignore the transverse extent 
of the string and imagine the string lives in a 1+1 dimensional space.  
As the $q\bar{q}$ pair oscillates, in one period it sweeps out an 
invariant area given by
\begin{equation}
	\Delta x_+ \Delta x_- = \frac{m^2}{\sigma^2}
	\label{eqn:invarientarea}
\end{equation}
in light-cone coordinates.  Here $m$ is the mass of the hadron and $\sigma$ is 
the string tension.  A hadron breaking off from a string is pictured in 
Fig.~\ref{fig:string}.

To assess any correlations between the mass and the production location of a 
hadron, we examine the distribution of break-up points of the string that 
produces the hadron.  In Fig.~\ref{fig:string}, points 1 and 2 are the 
locations where a $q\bar{q}$ pair separates from a string.  Assuming a 
constant break-up probability per unit time per unit length, $P$, the 
distribution of break-up points is given by
\begin{equation}
	d{\cal P}\propto \exp{(-Px^{(1)}_-x^{(2)}_+)}\theta{(x^{(1)}_--x^{(2)}_-)}
	\theta{(x^{(2)}_+-x^{(1)}_+)}.
	\label{eqn:breakup1}
\end{equation}
The exponential in this expression gives the probability that the string does 
not break up before the $q$ and $\bar{q}$ are made at points 1 and 2 and the 
theta functions ensure the proper ordering of the coordinates.  
We define the creation point of the hadron of interest as the mean of the 
break-off points of the $q\bar{q}$ pair:
\begin{equation}
	X_{\pm}=\frac{1}{2}(x^{(1)}_\pm + x^{(2)}_\pm).
\end{equation}
We also write 
\begin{equation}
	\Delta x_{\pm}=\pm(x^{(2)}_\pm - x^{(1)}_\pm).
\end{equation}
In terms of these, we may write Eq.~\eqref{eqn:breakup1} as
\begin{equation}\begin{array}{ll}
	\displaystyle
	     d{\cal P}\propto& \theta{(\Delta x_-)}\theta{(\Delta x_+)}\\
	&\displaystyle\times\exp\left(-P\left(X_{+}X_{-}
	   +\frac{1}{4}\Delta x_+\Delta x_-\right.\right. \\
	&\displaystyle\left.\left.+\frac{1}{2}(\Delta x_-X_{+}
           +\Delta x_+X_{-})\right)\right).
	\label{eqn:breakup2}
\end{array}\end{equation}
For a hadron at mid-rapidity, from Eq.~\eqref{eqn:invarientarea} we have 
$\Delta x_{+}=\Delta x_{-}=\frac{m}{\sqrt{2}\sigma}$.  Writing the hadron position 
back in Cartesian coordinates, we find 
\begin{equation}
	d{\cal P}\propto\exp\left(-\frac{P}{2}\left(T^2-X^2+
	   \frac{m}{\sigma}T\right)\right)\theta(T-|X|)
	\label{eqn:breakup3}
\end{equation}
This last theta function makes causality explicit.

Since $d{\cal P}$ expresses the probability of creating a hadron at position $X$ and
at time $T$, it is proportional to the emission rate in 
Eq.~\eqref{eqn:emissionrate}.  Thus, we can imagine doing the convolution in 
Eq.~\eqref{eqn:sourcefromrates} to obtain the source size.  Given that the emission
rates in this case are not Gaussian, we should not expect the source function itself to 
be Gaussian.  Nevertheless, we can still estimate the width of the source function 
from Eq.~\eqref{eqn:breakup3}.  We take the source width to be the distance at which
the magnitude of the source function drops by $1/e$.  From Eq.~\eqref{eqn:breakup3}, 
we see that the probability of creating a hadron drops by $1/e$ by roughly
\begin{equation}
	T\sim X\sim\frac{2\sigma}{mP}.
\end{equation}
This implies that the source function itself will have a width which is 
correlated with the mass of the hadron and given by
\begin{equation}
	R \sim \frac{2\sqrt{2} \sigma}{mP}.
	\label{eqn:stringsourcewidth}
\end{equation}
Here, the factor of $\sqrt{2}$ arises from the convolution in 
Eq.~\eqref{eqn:sourcefromrates}.  
If we make the reasonable assumption that the string tension and break-up probability 
are both universal, than we have the scaling relation in Eq.~\eqref{eqn:scaling}.  
Looking at the kaon source in Fig.~\ref{fig:NA44-sou}(d), the kaon source $1/e$-width 
is $R_K\approx 6$~fm.  Following the scaling, the proton source should drop by 
$1/e$ by roughly $R_p\approx \frac{m_K}{m_p} R_K=3$~fm.  This is roughly born 
out in the image in Fig~\ref{fig:NA44-sou}(f).  With 
\eqref{eqn:stringsourcewidth}, we can go further than than just 
checking this scaling and try to compute the source width directly. 
Taking typical values for the string break-up probability, $P=1$~fm$^2$, and for the
string tension, $\sigma=1$~GeV/fm, we find that the proton source radius should be about
R=2.9~fm and the kaon source radius should be about R=5.7~fm, again in rough agreement
with the images.  It would be very interesting to see if this scaling persists in 
sources imaged from other like-pair correlation in p-A collisions such as 
$\bar{p}\bar{p}$, $K_0K_0$ or $\Lambda\Lambda$.

In this consideration, we have neglected several things.  First, although 
we have ignored the transverse degrees of freedom of the strings, the longitudinal
length of the production region is much larger than the transverse extent so it
is the longitudinal extent that determines the angle-averaged source radius.
Second, we have ignored the finite mass of the quarks.  Adding finite quark masses
would change the trajectories of the $q\bar{q}$ pair and the shape of the area swept 
out by the pair, but it should not change our conclusions appreciably.
Third, we have neglected any consideration of other strings in the p-Pb interaction 
zone.  Next, if there are more than one string produced in the reaction, the possibility for
string fusion exists and it is neglected here.  Finally, we have neglected the 
Bjorken-like position-momentum correlations along the string length.  
These correlations will narrow the source~\cite{panitkin_99_1} and possibly account for 
the minor difference between the predicted and measured proton source widths above.

\section{Conclusion}

In this paper, we have investigated the possibility of detecting non-Gaussian 
sources in heavy-ion collisions.  In simple, but realistic, model calculations we
have demonstrated that it is possible to distinguish between Gaussian and 
non-Gaussian source shapes using an improved imaging method and high 
resolution data.  Imaging not only has achieved results comparable to Gaussian 
fits, but has now uncovered deviations from Gaussian behavior.

This improved imaging method has several features that make it superior to
the previous methods in~\cite{dbrown_1,dbrown_2}.  First, this method uses 
Basis splines to represent the source, giving a continuous 
representation of the source.  The resolution of the images is controlled by 
the placement of the knots (the points where the polynomials that make up the  
spline basis are matched).  Since the knots are analogous to the edges of the 
bins of a box spline representation, we may do as in~\cite{dbrown_2} and use 
the ``optimal knots'' to further improve the image.  In addition to these
improvements, equality constraints are now implemented in a simple manner.  
As in the previous imaging method, imaging is 
still a least-square problem in which the coefficients of the Basis 
spline representation are chosen to minimize the $\chi^2$
of the data.  In other words, the imaging gives the source that has the highest
probability of representing the correlation data.  Finally, we mention that
the amount of information available in the data limits the amount of information
we may extract in form of image and constraints ``increase'' the amount of
information.

Using this improved imaging method, we analyzed the proton, kaon, and pion 
correlations from S-Pb and p-Pb reactions measured by NA44.  We find evidence 
for non-Gaussian halos in the source functions from the like-meson correlations 
in the S-Pb reaction.  These halos are likely to be due to resonances 
decaying and producing the mesons.  
RQMD model simulations of the correlations reproduce 
the experimental data quite well except in the case of the proton correlations 
from the S-Pb reaction.  Unfortunately the source image does not shed much light 
on the discrepancy.  In the case of the p-Pb reaction, the imaged sources 
suggest a scaling of the source widths that we should expect on the basis of 
Lund-type string models.  This scaling should be tested by examining
other like-pair correlations (e.g. $\bar{p}\bar{p}$ and $\Lambda\Lambda$)
where these other pairs do not suffer from large resonance contributions.  

\section*{Acknowledgements}
We wish to thank Drs.~G.~F.~Bertsch, S.~Pratt, S.A.~Voloshin, N.~Xu, G.~Verde, 
and S.~Panitkin for their stimulating discussions.  We also want to give 
special thanks to Dr. Barbara Jacek for directing us to the NA44 correlations 
and to Dr. Giuseppe Verde for his careful reading of the manuscript.
This research was supported by National Science Foundation Grant No. 
PHY-0070818 and the U.S. Department of Energy Grant No. DOE-ER-41132.  
This work was performed under the auspices of the U.S. Department of Energy 
by the University of California, Lawrence Livermore National Laboratory under 
contract No. W-7405-Eng-48.

\nobreak
\appendix
\section{Basis Spline Representation of Sources}
\label{append:bsplines}

In this paper and previously in~\cite{dbrown_2}, we write the source expanded in some
basis:
\begin{equation}
	S(r)=\sum_{i=1}^{N_M} S_i B_i(r),
   \label{eqn:genericftnexpansion}
\end{equation}
In~\cite{dbrown_2} this basis was the box-spline basis.  In the box-spline basis, the
widths of the individual bins could be varied to increase or decrease the 
resolution of the kernel to minimize the relative error of the source. 
Unfortunately, in this representation the source is not continuous. 

In this work, we expand our sources in a more general basis:  
that of Basis splines, a.k.a. b-splines~\cite{de_boor}.  
B-splines are piece-wise continuous polynomials which can be made
arbitrarily smooth by changing the degree of the splines;  
the $0^{th}$ degree b-splines are actually box-splines used before.  
The b-splines are characterized by a set of knots which mark the points where
the various polynomials that make up the b-spline are matched.  In a sense,
these knots generalize the ``edges of the bins'' of the box-splines.
For this reason, we may vary the locations of the knots to minimize the 
relative error of the source, generalizing the method in~\cite{dbrown_2}.

\subsection{The basis splines}

Now we define the b-splines.  A $N_b^{th}$ degree b-spline  
is characterized by a set of knots, $\{t_k\}$ with $t_1\leq t_2 \leq ... \leq   
t_{N_{knots}}$.  The number of knots must be chosen so that 
$N_{knots}\geq N_M+N_b+1$ for $N_M$ b-splines.  For $N_b=0$, the 
b-splines are box splines, i.e.
\begin{equation}
   B_{0,i}(r)=X_i(r)\equiv\left\{
      \begin{array}{ll}
         1 &\mbox{if $t_i<r<t_{i+1}$}\\
         0 &\mbox{otherwise.}
      \end{array}
      \right.
\end{equation}
Note, if $t_i=t_{i+1}$ then $B_{0,i}(r)=0$.
The rest of the b-splines may be constructed from this first one through a 
set of recurrence relations
\begin{equation}
   \begin{array}{rl}
      B_{N_b+1,i}(r)=&\displaystyle w_{N_b+1,i}(r)B_{N_b,i}(r)\\
      &\displaystyle +(1-w_{N_b+1,i+1}(r))B_{N_b,i+1}(r)
   \end{array}
\end{equation}
where the weight factor is
\begin{equation}
   w_{N_b+1,i}(r)=\left\{
      \begin{array}{ll}
         \displaystyle \frac{r-t_i}{t_{i+N_b}-t_i} &
            \mbox{if $t_i\neq t_{i+N_b}$}\\
         0 &\mbox{otherwise.}
      \end{array}
      \right.
\end{equation}
With these definitions, we may write a b-spline of any degree back in terms of the 
box-splines:
\begin{equation}
	B_{N_b,i}(r)=\sum_{j=i}^{i+N_b-1} b_{N_b,j} X_j(r)
\end{equation}
where $b_{N_b,j}$ is a polynomial in $r$ of degree $N_b$ which we will not 
write explicitly.  We plot sample b-splines of different degrees in 
Fig.~\ref{fig:bsplines}.   

There are three other properties of the b-splines that are of note.  First, 
b-splines are normalized so that
\begin{equation}
	\int_{-\infty}^{\infty}\dn{}{r} B_i(r)=\frac{t_{i+N_b+1}-t_i}{N_b+1}.
\end{equation}
Second, the $i^{th}$ b-spline is zero outside of the region 
$t_i\leq r\leq t_{i+N_b+1}$.  Third, the b-splines are not orthogonal but
expressions for their inner product exist~\cite{de_boor}.

From the definitions and from the figure, it is not clear how to pick the knots.
The knots do not have to be equally spaced and, in many situations, it is best
{\em not} to space them equally.  In fact, one can even pile up to $N_b+1$ 
knots on the same point!  One can do this because
\begin{center}
   $N_b+1$ = number knots at a point + number continuity conditions at that point.
\end{center}
If a b-spline has $N_b+1$ knots at a point then that b-spline is discontinuous
there.  Away from these multiple knots, the b-spline is still continuous up to
derivatives of degree $N_b$.  In the main text, we remove all assumptions about the
continuity of the source at the boundaries by keeping $N_b+1$
knots at the boundaries.  We then optionally re-insert continuity conditions using 
equality constraints.

\subsection{Optimizing resolution}

Now we come to the point of choosing the best knots for the inversion,
generalizing the ``optimized discretization'' 
scheme of Ref.~\cite{dbrown_2}.  First, the model covariance 
matrix (cf. Eq.~\eqref{eqn:modelcovmtx}) depends on the kernel of the inversion, 
the error on the data and whatever scheme we use to represent the source, 
{\em but not on the data or the model itself}.  For a given
kernel and set of data errors, we are free to change our representation of the
source in order to minimize the error of the source.  In particular, we may 
vary the location of the knots (at least not the knots fixed at the endpoints 
of the imaging region) to minimize the error of the source coefficients,
$\Delta S_{j}=\sqrt{\Delta^2S_{jj}}$, relative to some dummy source: 
\begin{equation}
   \sum_{j=N_b+2}^{N_{knots}-N_b-1}\left|\frac{\Delta S_{j}}
   {S^{dummy}_j}\right|=\mbox{min}.
\end{equation}
The coefficients $S^{dummy}_j$ are the coefficients of the 
expansion of a dummy source in b-splines.
In this minimization, the first and last $N_b+1$ knots are held fixed and the
positions of all of the other knots are varied.  The dummy source itself is 
chosen to be big roughly where one expects the source to be big and small where 
one expects the source to be small.  Since the detailed shape of the 
dummy source should not be important, in this paper we chose an exponential 
dummy source with radius $R^{dummy}=3.5$~fm given by 
$S^{dummy}(r)\propto\exp{(-r/R^{dummy})}$.

\section{Bayesian Approach to Imaging}
\label{append:imaging}

In this appendix, we will explain the technical details of the Bayesian Approach
to imaging and extend the approach of Ref.~\cite{dbrown_2} 
by implementing constraints in a more consistent manner.
In the previous works the constraints are implemented by Monte-Carlo   
sampling the experimental errors, leading to statistical fluctuations in the 
extracted source.  In that approach, no distinction is made between equality and
inequality constraints. By equality constraints, we mean constraints of the form
\begin{equation}
	\int \dn{}{{\bf r}} f({\bf r}) S_{\bf P} ({\bf r})= \mbox{constant},
   \label{eqn:genericeqcon}
\end{equation}
and by inequality constraints, we mean constraints of the form
\begin{equation}
	\int \dn{}{{\bf r}} f({\bf r}) S_{\bf P} ({\bf r})\geq\mbox{constant}.
   \label{eqn:genericineqcon}
\end{equation} 
Both of these types of constraints are forms of {\em prior information}, meaning
information we have in hand before we began imaging.  In this appendix, 
we will explain how to use prior information in an inversion.  In
particular, we will discuss two methods for implementing equality 
constraints {\em directly} into the inversion and we will discuss a better 
way to implement inequality constraints.  We conclude this appendix with a
brief discussion of inequality constraints.

\subsection{General Theory}

Suppose we have an $N_D$ dimensional vector of observed data ${\bf d}^{obs}$ 
with covariance matrix $\Delta^2d$ that we wish to represent by some $N_M$ 
dimensional model vector ${\bf m}$.  In the case of source imaging, our model 
vector is the vector of coefficients of the function expansion of the source and
the data is the raw correlation function.  We assume the data has a diagonal 
covariance matrix.  In an ideal measurement of the data, there would be no 
experimental uncertainty and the data and the model would be related through 
some linear equation:
\begin{equation}
	{\bf d}=K\cdot{\bf m}.
\end{equation}
Here, $K$ is the kernel of an integral equation.

In a real imaging problem, the observed data has errors and
statistical scatter.  To make progress, we adopt the
so-called Bayesian Approach to imaging where we seek the probability density, 
$\sigma({\bf m})$, for a specific model ${\bf m}$ to represent the
data~\cite{tarantola,Gouveia97}.  With this density, we take the mean of the
density as an estimator of the true model and the width of the density as an
estimate of the uncertainty in our model.  Neglecting the error in our
determination of the kernel and assuming the uncertainty in our measurement of 
${\bf d}^{obs}$ is Gaussian, we can write Bayes Theorem as follows:
\begin{equation}
	\sigma({\bf m})\propto\rho({\bf m})\exp\left(-\frac{1}{2}\chi^2_{data}\right).
\end{equation}
Here, $\rho({\bf m})$  encodes all of the prior information we have about the 
model and $\chi^2_{data}$ is:
\begin{equation}
	\chi^2_{data}=\left(K\cdot{\bf m}-{\bf d}^{obs}\right)^T\cdot(\Delta^2d)^{-1}\cdot
		\left(K\cdot{\bf m}-{\bf d}^{obs}\right).
\end{equation}
Here the superscript $T$ represents a matrix transpose.  The 
dimension of the model vector, $N_M$, and the dimension of the data vector,
$N_D$, need not be equal.  Indeed, it is better to have many more data points 
than model parameters so that we may over-constrain the system.
For the time being, assume we have no prior information so we may set 
$\rho({\bf m})=1$.

We immediately see that the most probable model vector is the one that
maximizes the probability and hence minimizes the $\chi^2_{data}$.  Following 
\cite{dbrown_1,dbrown_2,tarantola,parker,NumRecipes,Gouveia97}, we can find 
both the model vector that minimizes this $\chi^2_{data}$ as well as the 
covariance matrix of the model:
\begin{equation}
	\left<{\bf m}\right>=\Delta^2m\cdot 
	K^T\cdot(\Delta^2d)^{-1}\cdot {\bf d}^{obs}
	\label{normal_eqn}
\end{equation}
and
\begin{equation}
	\Delta^2m=(K^T\cdot(\Delta^2d)^{-1}\cdot K)^{-1}
   \label{eqn:modelcovmtx}.
\end{equation}
Eq.~\eqref{normal_eqn} usually goes by the name of a normal 
equation.  The model covariance matrix is independent of ${\bf d^{obs}}$ and 
depends only on the error of the data and the kernel itself.  

We can write the $\chi^2_{data}$ directly in terms of the 
model covariance matrix and the $\chi^2_{data}$ minimizing model vector~\cite{tarantola}:
\begin{equation}
	\chi^2_{data}=\left({\bf m}-\left<{\bf m}\right>\right)^T\cdot 
		(\Delta^2m)^{-1}\cdot\left({\bf m}-\left<{\bf m}\right>\right).
\end{equation}
In other words, if we have a $N_M$ dimensional model space, then
the $\chi^2_{data}=1$ hypersurface is an $N_M$-dimensional hyper-ellipsoid with 
principal axes given by the eigenvectors of the covariance matrix of the model, 
$\Delta^2m$.  These principal axes do not need to correspond to
the directions corresponding to the $m_j$ components in the model space.

\subsection{Equality Constraints}

Now let us discuss the role of prior information in the general inversion 
problem.  We concentrate equality
constraints such as in Eq.~\eqref{eqn:genericeqcon}.
In a matrix form, equality constraints are written as
\begin{equation}
   {\cal C}\cdot{\bf m}={\bf c}.
   \label{equalitycon}
\end{equation}
Here ${\cal C}$ is a matrix of constraint equations and ${\bf c}$
is a constant vector of constraint values. 
In the inversion problem in the main text, there are a variety of 
constraints we might use and they are listed in Table~\ref{table:eqcons}.
The equality constraint in~\eqref{equalitycon}
corresponds to the prior probability density of 
\begin{equation}
	 \rho({\bf m})\propto\delta\left({\cal C}\cdot{\bf m}-{\bf c}\right).
\end{equation}
Equality constraints can be cast into a Gaussian prior probability density 
simply by writing this density as a Gaussian with vanishing width:
\begin{equation}
	\rho({\bf m})\propto\lim_{\lambda\rightarrow\infty}
	\exp{\left(-\frac{\lambda}{2}\chi_{equal}^2\right)}	
\end{equation}
where
\begin{equation}
   \chi_{equal}^2=\left({\cal C}\cdot{\bf m}-{\bf c}\right)^2.
   \label{eqn:equalitychi}
\end{equation}
Finding the most probable model then corresponds to minimizing a modified 
${\tilde{\chi}}^2$:
\begin{equation}
    {\tilde{\chi}}^2=\chi^2_{data}+\lambda\chi^2_{equal}.  
   \label{modified_chi_square}
\end{equation}
The solution is straightforward and corresponds to the most probable model:
\begin{equation}
	   \left<{\bf m}\right>=
      \Delta^2m\cdot\left(K^T\cdot(\Delta^2d)^{-1}\cdot {\bf d}^{obs}
      +\lambda{\cal C}^T\cdot{\bf c}\right)
   \label{really_new_normal_eqn}
\end{equation}
along with the model covariance matrix:
\begin{equation}
	\Delta^2m=
   \left(K^T\cdot(\Delta^2d)^{-1}\cdot K
   +\lambda{\cal C}^T\cdot{\cal C}\right)^{-1}.
   \label{really_new_covarmtx}
\end{equation}

It is clear that to correctly simulate a delta function prior information
density, we must choose a large $\lambda$.  Looking at
Eq.~\eqref{modified_chi_square}, we must choose a $\lambda$ so that
$\lambda\chi^2_{equal}\gg\chi^2_{data}$.  We now estimate the sizes of 
$\chi^2_{equal}$ and $\chi^2_{data}$.  A good fit to the data should have the 
$\chi^2_{data}$ nearly at the number of degrees of freedom, i.e. 
$\chi^2_{data}\approx N_D-N_M$.  To estimate $\chi^2_{equal}$, we follow 
Eq.~\eqref{eqn:equalitychi}.  In the main text, a typical source is 
$\gtrsim 1\times 10^{-6}$ fm$^{-3}$ and the constraint matrix 
and vectors are typically $\sim 1$ fm$^3$, so 
$\chi^2_{equal}\sim N_M\times 10^{-12}$.
Putting this together,  we must choose 
\begin{equation}
	\lambda\gg \chi^2_{data}/\chi^2_{equal}\sim 
	\left(\frac{N_D}{N_M}-1\right)\times 10^{12}.
\end{equation}
For example, for $N_D=83$ and $N_M=8$ we need $\lambda\gg 10^{11}$.
By adjusting strength of $\lambda$, we can adjust strength of 
various terms, emphasizing stability of inversion (i.e., obeying constraints) 
over representing the data.  Thus, $\lambda$ functions as a trade-off parameter
in the jargon of inverse theory. See section 18.4 of {\em Numerical 
Recipes}~\cite{NumRecipes} for a more complete discussion.  

A useful alternative to this scheme (and a way to do 
$\lambda\rightarrow\infty$ limit exactly) is to use the Householder 
transformation to eliminate the constraints from the unmodified normal equations
of Eqs.~\eqref{normal_eqn} 
\cite{NumRecipes,cernlib}.  
The trade-off is that the Householder transformation may be somewhat unforgiving. 
Due to an unfortunate choice of basis functions, it may not be possible to
satisfy two constraints simultaneously even if they can be satisfied 
simultaneously in the true answer.  By keeping $\lambda$ finite, we are never
trying to satisfy the constraints exactly so can do a reasonable job of obeying
both constraints.  Nevertheless, schemes based on 
Householder reductions of the constraints are complementary to ones using the 
Gaussian prior probability.

\subsection{Inequality Constraints}

Now we ask how to use constraints of the form 
\begin{equation}
  {\cal C}\cdot{\bf m}\geq {\bf c}.
\end{equation}
Such constraints are called inequality constraints and there are many different
ones we could use:  the source is positive definite, the derivative
of the source is bounded (to ensure smoothness), or the source satisfies the
Fourier transform test from Ref.~\cite{Brown:1999ka}.
Inequality constraints correspond to prior probability densities of the form
\begin{equation}
	\rho({\bf m})\propto\theta\left({\cal C}\cdot{\bf m}-{\bf c}\right)
\end{equation}
which cannot be rendered into a Gaussian form.

In Ref.~\cite{dbrown_1}, we use a simple Monte-Carlo sampling scheme 
to implement inequality constraints.  In this scheme, 
one uses the experimental uncertainty to generate an ensemble of correlations,
each consistent with the original.  One then inverts each one to obtain a sample
source and discards any sources that are not consistent with the inequality
constraints.  One then combines the samples that are consistent with the
constraints to obtain an average source and an estimate of the errors on the
source.  The problem with this scheme is that it pushes the sources away
from edges of the model space defined by the constraints.

\begin{figure}
	\epsfig{file=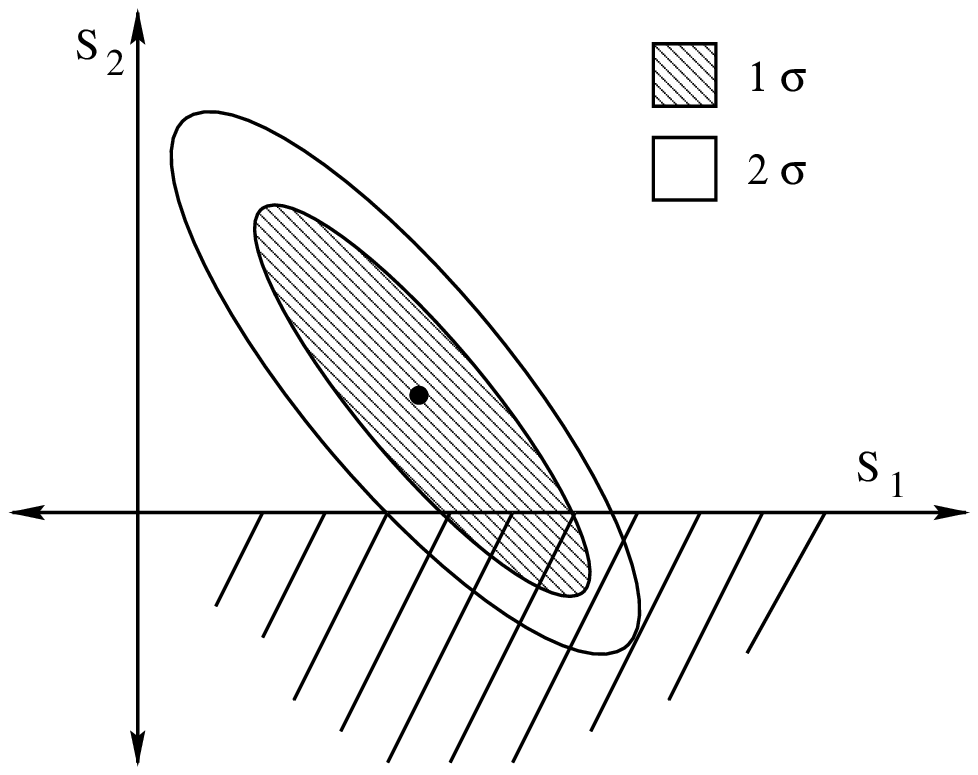,width=0.45\textwidth}
	\caption{An illustration of an inequality constraint cutting
			through 1$\sigma$ band of a best-fit region.}
   \label{fig:ineq}
\end{figure}

We illustrate this problem with a simple example.  Suppose we have an inversion
problem where the goal is to determine two points $S_1$ and $S_2$ under the 
constraint that $S_2>0$.  We sketch one possible outcome of the inversion in
Fig.~\ref{fig:ineq}.  In this picture, we see the best fit value of $S_1$ and
$S_2$ is consistent with our inequality constraint, but the constraint cuts
through both the $1\sigma$ and $2\sigma$ contours.  Using the Monte-Carlo
sampling scheme discussed above, we would actually be finding a false best-fit
point which is slightly above and to the left of the true
best-fit value because we throw out samples with $S_2<0$.  
The errors on these points would also be symmetrically placed
around this point.  In fact the correct way to solve the problem is just to 
quote the best-fit values of $S_1$ and $S_2$, with asymmetrical errors.

The way inequality constraints are implemented in most commercial inversion
packages is through so-called ``Active Set Methods''~\cite{NEOSGuide}.  In 
these methods, one finds the best-fit solution as one normally would have if 
there were
no inequality constraints.  If the best-fit solution lies in a region excluded
by the inequality constraints, then the code finds the edge of the included
region (the so-called Active Set) and searches along it until the code finds the
solution that minimizes the $\chi^2$.  Such a scheme is powerful, but likely
beyond what is needed for our problem.


\pagebreak
\onecolumn
\begin{center}TABLES\end{center}
%
%
\begin{table}
   \begin{tabular}{|c|c|c|}
      constraint & continuous representation & b-spline representation \\
      \hline
      flat at $r=0$ &
      $\displaystyle\frac{\partial S}{\partial r}(r\rightarrow 0)=0$&
      $\displaystyle\sum_{j=1}^{N_M} S_j B_j'(r\rightarrow 0)=0$\\
      \hline
      normalized to $\lambda$ &
      $\displaystyle 4\pi\int_0^\infty \dn{}{r}r^2 S(r)=\lambda$&
      $\displaystyle\sum_{j=1}^{N_M} 4\pi S_j \int_0^\infty \dn{}{r}r^2
      B_j(r)=\lambda$\\
      \hline 
      zero outside of imaged region &
      $\displaystyle S(r_{max})=0$&
      $\displaystyle \sum_{j=1}^{N_M} S_j B_j(r_{max})=0$\\
      \hline
      flat at $r=r_{max}$&
      $\displaystyle \frac{\partial S}{\partial r}(r_{max})=0$&
      $\displaystyle \sum_{j=1}^{N_M} S_j B_j'(r_{max})=0$\\
   \end{tabular}
   \caption{Equality constraints on the b-spline representation of spherically
      symmetric sources.}
   \label{table:eqcons}
\end{table}
%
%
\begin{table}
   \begin{tabular}{rrccccc}
          &&$r_{max}$ [fm]&$\#$ coeffs& $r=0$ constraint?&$\#$ data pts.&$\chi^2$\\
      \hline
      S-Pb&$\pi^+\pi^+$ &35&7&yes&29(7)&19.8\\
          &$K^+K^+$     &35&7&yes&16(8)&5.0\\
          &$pp$         &26&6&yes&20(6)&14.6\\
      \hline
      p-Pb&$\pi^+\pi^+$ &21&5&yes&29(9)&24.8\\
          &$K^+K^+$     &26&8&yes&29(9)&23.1\\
          &$pp$         &26&8&no&20(8)&7.6\\
  \end{tabular}
\caption{Parameters used in the reconstruction of the NA44 sources.  The
numbers in parentheses in the number of data points column is our estimate of
the number of points which contain usable information.}
\label{table:imag_param}
\end{table}
%
%
\begin{table}
   \begin{tabular}{ccr@{$\pm$}lr@{$\pm$}lr@{/}l}
      & Ref. &\multicolumn{2}{c}{$\lambda$} & \multicolumn{2}{c}{$R_{G}$} [fm]&
         \multicolumn{2}{c}{$\chi^2/NDF$}\\
      \hline
      $K^+K^+$ (S-Pb) &\protect\cite{NA44_K}&0.92&0.08&3.22&0.20&53&31\\
      \hline
      $\pi^+\pi^+$ (S-Pb)&\protect\cite{NA44_pi_1}&0.46&0.04&4.50&0.31&18.1&16\\
      \hline
      $K^+K^+$ (p-Pb) &\protect\cite{NA44_K}&0.68&0.06&1.71&0.17&65&54\\
      \hline
      $\pi^+\pi^+$ (p-Pb) &\protect\cite{NA44_pi_1}&0.38&0.03&2.89&0.30&16&25\\
  \end{tabular}
\caption{Gaussian fit parameters as obtained by NA44.}
\label{table:gauss_fit_param}
\end{table}
%
%
\begin{table}
   \begin{tabular}{ccr@{$\pm$}lr@{$\pm$}lr@{/}l}
      & Ref. &\multicolumn{2}{c}{$\lambda$} & \multicolumn{2}{c}{$R_{D}$} [fm]&
         \multicolumn{2}{c}{$\chi^2/NDF$}\\
      \hline
      $K^+K^+$ (S-Pb) &\protect\cite{NA44_K}&1.80&0.18&2.64&0.22&26&31\\
      \hline
      $\pi^+\pi^+$ (S-Pb) &\protect\cite{NA44_pi_1}&0.77&0.08&3.54&0.33&12.0&16\\
      \hline
      $K^+K^+$ (p-Pb) &\protect\cite{NA44_K}&1.10&0.10&1.04&0.19&55&54\\
  \end{tabular}
\caption{Exponential fit parameters as obtained by NA44.}
\label{table:exp_fit_param}
\end{table}
%
%
\begin{table}
   \begin{tabular}{c c r@{$\pm$}l r@{$\pm$}l r@{$\pm$}l r@{$\pm$}l}
      & {$\left<p_T\right>$ [MeV/c]}
      & \multicolumn{2}{c}{$\lambda$} 
      & \multicolumn{2}{c}{$R_{Ts}$} [fm]
      & \multicolumn{2}{c}{$R_{To}$} [fm]
      & \multicolumn{2}{c}{$R_{L}$} [fm]\\
      \hline
      $\pi^+\pi^+$  &150 &0.56&0.02 &4.15&0.27 &4.02&0.14 &4.73&0.26 \\
      \hline
      $\pi^+\pi^+$  &450 &0.55&0.02 &2.95&0.16 &2.97&0.16 &3.09&0.19 \\
      \hline
      $K^+K^+$      &240 &0.82&0.14 &2.55&0.20 &2.77&0.12 &3.02&0.20 \\
      \hline
  \end{tabular}
\caption{Three dimensional Gaussian fit parameters from the S-Pb reaction 
as obtained by NA44.}
\label{table:gauss3D_fit_param}
\end{table}
\pagebreak
\onecolumn
\begin{center}FIGURES\end{center}
%
%
\begin{figure}
   \epsfig{file=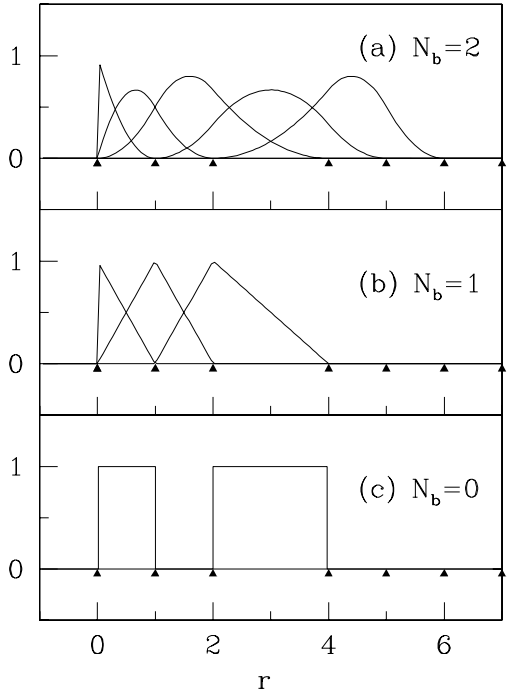,angle=0,width=0.40\textwidth}
   \caption{Sample plots of $N_b^{th}$ degree b-splines.  
   In all panels, the knots are marked by carets and the knots at 
   $r=0$ are actually $N_b+1$ regular knots piled together.}
   \label{fig:bsplines}
\end{figure}
%
%
%
%

%
%

%
%

%
%

%
%

%
%

%
%
%
%

%
%

%
%

%
%

%
%
%
%
\pagebreak

%
%

\pagebreak
%
%

\end{document}